\documentclass{aa}
\usepackage{txfonts}
\usepackage{graphicx}
\usepackage{mathtools}
\usepackage{ulem}
\usepackage{orcidlink} 
\DeclareUnicodeCharacter{0327}{\c{}}

\makeatletter
\renewcommand*\aa@pageof{, page \thepage{} of \pageref*{LastPage}}
\makeatother

\begin{document}

\title{From Streaming Instability to the Onset of Pebble Accretion}
\subtitle{I. Investigating the Growth Modes in Planetesimal Rings}

\author{Nicolas Kaufmann\inst{1}, Octavio M. Guilera \orcidlink{0000-0001-8577-9532} \inst{2},Yann Alibert\inst{1} \and Irina L. San Sebasti\'an\inst{2,3}}

\institute{Physikalisches Institut, Universität Bern, Gesselschaftsstrasse 6, 3012 Bern, Switzerland,
\and
Instituto de Astrof\'{\i}sica de La Plata, CCT La Plata-CONICET-UNLP, Paseo del Bosque S/N (1900), La Plata, Argentina,
\and
{Department of Aerospace Science and Technology, Politecnico di Milano, Milano, 20156, Italy.}
}
\date{\today}

\abstract{The localized formation of planetesimals can be triggered with the help of streaming instability when the local pebble density is high. This can happen at various locations in the disk leading to the formation of local planetesimal rings. The planetesimals in these rings subsequently grow from mutual collisions and by pebble accretion.}
{We investigate the early growth of protoplanetary embryos from a ring of planetesimals created from streaming instability to see if they reach sizes where they accrete pebbles efficiently.}
{We simulate the early stages of planet formation for rings of planetesimals that we assume were created by streaming instability at various separations from the star and for various stellar masses using a semi-analytic model.}{The rings in the inner disk are able to produce protoplanetary embryos in a short time whereas at large separations there is little to no growth. The growth of the largest bodies is significantly slower around lower-mass stars. }
{The formation of planetary embryos from filaments during the disk lifetime is possible but strongly dependent on the separation from the star and the mass of the host star. It remains difficult to form the seeds of pebble accretion early in the outer disk $\sim 50AU$, especially for low-mass stars.}


\authorrunning{Kaufmann et al.}
  
\maketitle

\section{Introduction}

In order to form planets in protoplanetary disks the initially micron-sized dust has to grow many orders in magnitude. On this size ladder, there are many barriers to the growth through coagulation, for example, the fragmentation barrier or the limitation of growth through radial drift as the dust decouples from the gas. This typically halts the growth of the dust at pebble-sizes ($\sim$mm/cm) defined by their Stokes number $\tau_s\approx 10^{-2}$--$10^{-1}$\citep{guttler_outcome_2010,zsom_outcome_2010,krijt_erosion_2015,birnstiel_dust_2011,stammler_dustpy_2022, birnstiel_dust_2023}.

As a potential avenue to grow past this barrier, the streaming instability (SI) helps to concentrate the pebbles into clumps \citep{youdin_streaming_2005,johansen_protoplanetary_2007}. Then the direct gravitational collapse of these over-dense clumps of pebbles into planetesimals is a promising way to overcome this growth barrier. The Initial Mass Function (IMF) of planetesimals formed through these mechanisms is typically sharply peaked at around 100 kilometres \citep{johansen_zonal_2009,bai_dynamics_2010}. Still, the distribution strongly depends on the local properties of the gas disk and the properties of the collapsing pebbles \citep{schafer_initial_2017,klahr_turbulence_2020,polak_high_2022} and it is still poorly constrained. 

For the streaming instability to trigger we need a highly enhanced local dust-to-gas ratio in the midplane i.e. $\rho_{peb} \approx \rho_{gas}$. Various viable mechanisms could lead to such an enhancement of the dust-to-gas ratio at specific locations in the disk. The ice lines of different elements and molecules serve as such potential locations. For example, the silicate and water condensation lines have been investigated in this context as they account for a large portion of the mass budget thus leading to a bigger enhancement of the local dust-to-gas ratio. This happens due to the re-condensation of vapour outside of the respective ice line \citep{stevenson_rapid_1988,drazkowska_planetesimal_2017,schoonenberg_planetesimal_2017,abod_mass_2019, schneider_how_2021}. Further mechanisms to enhance the dust-to-gas ratio have been proposed including the dead zone inner edge where a transition in turbulence generates a pressure bump and can trap the drifting pebbles \citep{chatterjee_inside-out_2013, guilera_giant_2020} and the edge of a gap in the gas disk carved by giant planets is also able to accumulate pebbles leading to the formation of planetesimals \citep{shibaike_planetesimal_2020,shibaike_planetesimal_2023,sandor_planetesimal_2024}. Most of these formation mechanisms lead to the formation of planetesimals in small regions all over the disk which is in contrast to the classical picture of core accretion that usually starts with planetesimals distributed in the entire disk \citep{emsenhuber_new_2021,chambers_planet_2006,pollack_formation_1996}.

If there exist variations in the pressure gradient of the gas disk i.e. pressure bumps, regardless of their physical origin, they are also able to concentrate the drifting pebbles, thus enhancing the local dust-to-gas ratio and triggering the streaming instability. The formation of planetesimals and planets in these postulated structures has been investigated in many recent works \citep{jiang_survival_2021,stammler_dsharp_2019,lau_rapid_2022, sandor_planetesimal_2024}. Although the structures in the simulations are not modelled physically self-consistently, there exists ample observational evidence for their existence in observations \citep{bae_structured_2023,pinte_kinematic_2022}. In these pressure structures, the largest planetesimals formed from direct collapse are directly able to accrete pebbles efficiently with negligible contributions from mutual collisions among planetesimals. In these environments, the longer encounter times between the planetesimals and pebbles lead to a higher accretion efficiency. This happens because the radial drift of pebbles is severely slowed down as the headwind from the gas disk vanishes.

The formation of planetesimals in disks without structures has also been investigated in various works \citep{lenz_planetesimal_2019,voelkel_impact_2020,lau_can_2024,jiang_survival_2021}. These formation models are motivated by transient pebble traps that could trigger the streaming instability anywhere in the disk while not affecting the global disk evolution. The subsequent growth in such planetesimal filaments/rings in disks without structures has been studied by \cite{liu_growth_2019,jang_growth_2022} and \cite{lorek_growing_2022}. For these environments, the largest planetesimals formed are typically in the mass range where they accrete pebbles in the Bondi regime which typically leads to slow growth. The second source of accretion is from the mutual collisions of planetesimals that tends to be slow for large planetesimals as they are weakly coupled to the gas but may contribute significantly in these early stages \citep{liu_growth_2019}. Note that the growth in planetary rings differs from the classical runaway and oligarchic regime due to the fact that the width of the ring is typically smaller than the classical feeding zone so the maximal mass the largest body can reach is limited by the total mass of the filament rather than the classical isolation mass. However, most planet formation models based on pebble accretion insert larger embryos as initial conditions that can accrete pebbles efficiently \citep{lambrechts_forming_2014, guilera_importance_2021, drazkowska_how_2021}, to form planets during the lifetime of the gas disk. This timescale constraint is crucial for giant planets as they are required to form early enough before the gas disk dissipates to accrete significant gaseous envelopes. Therefore we investigate the growth in such planetesimal rings in unstructured disks to see if and on what timescale protoplanetary embryos can form. To do this we will simulate the growth in such filaments at different locations in the disk using our newly developed model. 
This study, although similar in setup, is complementary to the aforementioned studies and differs from their approaches in the following ways. In the works of \cite{liu_growth_2019} and \cite{jang_growth_2022}, they use N-body methods to describe the planetesimals which forces them to truncate the initial size distribution and treat planetesimal collisions as perfect mergers to have reasonable computation times. Whereas we use an Eulerian approach for our planetesimals allowing us to describe a larger number of planetesimals. In addition to the effects considered in the work of \cite{lorek_growing_2022} that investigates the growth in these filaments from planetesimal accretion we also consider the accretion of pebbles onto the embryo. Additionally, in our model, the distribution of planetesimals evolves in time by solving the continuity equation as opposed to their local approach. Lastly, we consider the full-size distribution when we calculate the fragmentation of planetesimals as opposed to a fixed size of fragments allowing us to more accurately capture the size evolution of the planetesimals and the associated accretional processes.

This paper is structured in the following way: In Section \ref{sec:Model} we present the planet formation model used to simulate the early growth of protoplanetary embryos from a ring of planetesimals including pebble accretion. In Section \ref{sec:Results} we show the results of our investigation and discuss how the different model parameters influence our results, explore the formation around stars of different masses and discuss the implications of our simulation for the timing of core formation. In Section \ref{sec:conlusions} we discuss our results, go over the limitations of our approach and give a final summary.

\section{Planet formation model and setup}
\label{sec:Model}
We model the formation of planetary embryos from an initial ring of planetesimals at different locations in the disk. This localised formation of a ring of planetesimal is motivated by the streaming instability \citep{youdin_streaming_2005,johansen_rapid_2007,schafer_initial_2017} that requires a local pebble over density to be triggered. This leads to the formation of a ring of planetesimals with a width that is dictated by the pressure gradient of the gas $\eta = -0.5 h^2 \frac{d\text{ln}P}{d\text{ln}r}$ disk and the separation of the star $r_0$ as $\Delta w \sim \eta*r_0$ \citep{yang_concentrating_2017,li_demographics_2019}. There are many possible mechanisms that can lead to the required concentration of pebbles for the streaming instability to happen. For example, the ice line \citep{drazkowska_planetesimal_2017} where the outwards diffusion of the water vapour leads to an enhanced enrichment of the disk. Other proposed mechanisms to concentrate the dust include the dead zone inner boundary \citep{hu_inside-out_2018}, differential drift due to the back reaction of the dust \citep{jiang_survival_2021} and the external photo-evaporation in the outer disk \citep{carrera_planetesimal_2017} or the gap opened by the other planets in the disk \citep{shibaike_planetesimal_2020,shibaike_planetesimal_2023}. 

Therefore we test the growth of a single discrete embryo in these planetesimal rings at different locations in a \textbf{smooth} disk without invoking a specific mechanism or structure for the formation. We use a semi-analytic model based on the \textbf{Bern} model \citep{emsenhuber_new_2021} that simulates the growth of a single discrete embryo that represents the single largest planetesimal formed in a ring of planetesimals with a size distribution. Additionally, we calculate the evolution of the gas disk and the pebble flux from the outer disk. First, we explain the evolution of the solids the accretion description used then we go over the disk model and finally explain the initial conditions used. 

\subsection{Evolution of the solids}
The solids in the disk are described by a two-component approach: the planetary embryo and the planetesimals. The planetary embryo is described as a discrete body of mass $M_{em}$ and accretes planetesimals as described in \cite{emsenhuber_new_2021}; \cite{chambers_semi-analytic_2006} and \cite{inaba_high-accuracy_2001}. Since the width of the initial filament is significantly smaller than the classical feeding zone of the embryo in the oligarchic regime, we have to define the mean local surface density of planetesimals used to calculate the accretion rate,
\begin{equation}
    \Dot{M}_{em} = h_{em,k}^2 r_0^2 \Omega_k \Sigma_k P_{col},
\end{equation}
where $h_{em,k} = \left(\frac{m_{em}+m_k}{3 M_\star}\right)^{1/3}$ and $P_{col}$ is the intrinsic collision probability according to \cite{inaba_high-accuracy_2001} and $\Sigma_k$ is the local surface density of the planetesimals of population $k$. To do this we calculate the mean surface density in an annulus around the embryo that contains $\text{erf}(2^{-0.5}) \approx 68\%$ of the total mass of the planetesimals. This choice was motivated by comparing our results with the analytical formula for the spreading of a planetesimal belt described in Eq. (17) of \cite{liu_growth_2019}. 

As this study focuses on the early stages of core formation the gas accretion of a planetary envelope is neglected. The core is assumed to have the same initial bulk density as the planetesimals $\rho_s$ and is considered to be on a circular orbit as the damping by dynamical friction of embryo from the planetesimals leads to a rapid decay of its eccentricity and inclination \citep{lorek_growing_2022}.

The planetesimals follow a fluid-like description on a grid and are characterized by their surface density $\Sigma$, their mean root squared eccentricity ($e$) and inclination ($i$) and their bulk density ($\rho_s$). They are described on a grid both as a function of the distance from the star ($a_i$) and planetesimal radius $r_{p_i}$, $\Sigma_p(a_i,r_{p_i})$ where the planetesimals occupy different logarithmic spaced bins $r_i$ according to their size, 
\begin{equation}
    r_{p_i} = (r_{max}/r_{min})^{(i-1)/(N-1)}\times r_{min},
\end{equation}
where $r_{p_i}$ is the size of the i'th planetesimal bin/population, $N$ refers to the number of size bins and $r_{max}$ and $r_{min}$ refer to the maximal and minimal size of planetesimals considered in the code. The surface density $\Sigma(r_{p_j})$ then describes mass contained in bodies between $[\sqrt{r_{p_i}*r_{p_{i+1}}},\sqrt{r_{p_i}*r_{p_{i-1}}}]$. The maximal size is dictated by the initial conditions described in Section \ref{sec:initial} and the minimal size is chosen to be $1$~cm. The dynamical state of each size bin of planetesimals is described by their mean squared eccentricity ($e$) and inclination ($i$) that are calculated by solving their evolution equation as described in \cite{kaufmann_influence_2023}. The model takes into consideration the damping by gas drag \citep{chambers_planet_2006} the stirring by other planetesimals \citep{ohtsuki_evolution_2002} and the embryo including dynamical friction \citep{emsenhuber_new_2021} and solves the evolution equation for $e$ and $i$. Additionally, we solve the continuity equation for the planetesimals given by 
\begin{align}
\label{eq:cont_solids}
    \frac{\partial}{\partial t}(\Sigma_i) &- \frac{1}{r}\frac{\partial}{\partial r}(r v_{drift}\Sigma_i) - \frac{1}{r}\frac{\partial}{\partial r} \left[ 3r^{0.5}\frac{\partial}{\partial r}(r^{0.5} \nu \Sigma_i)\right] \\
    &= \dot\Sigma_{accretion} + \dot\Sigma_{frag}\nonumber
\end{align}
where the drift is caused by the headwind experienced by the planetesimals due to the sub-Keplerian orbital velocity of the gas \citep{guilera_planetesimal_2014} however as the large planetesimals are only weakly bound to the gas we ignore the effects of radial drift in this study. The diffusion of the planetesimals due to their mutual gravitational interaction follows the description of \citet{ohtsuki_radial_2003}, and \citet{tanaka_new_2003}. However, this description of the viscosity can lead to a negative diffusion coefficient for small values of $\beta = i/e$ when it is significantly below the equilibrium value of $\beta = 0.5$ which may not be physical. This can occur in the zones where the stirring by the protoplanetary embryo falls off and in the early stages of formation there the stirring rates for the eccentricities are significantly higher than for the inclinations \citep{liu_growth_2019,ohtsuki_evolution_2002}. For simplicity, we consider the averaged diffusion rate over the entire filament by calculating the viscosity using the average surface density in the filament and the values of $e$ and $i$ at the embryos's location.

\subsection{Planetesimal fragmentation model}
\label{sec:frag_theory}

To investigate the growth of the embryo in these filaments we have to consider the fragmentation of planetesimals due to mutual collisions among them as it can have an impact on the growth timescales involved due to the evolution of the size distribution of the planetesimals \citep{lorek_growing_2022}. 

The outcome of the collisions among planetesimals is determined by both their material strength and the kinetic energy involved in the collisions. The mass distribution of the bodies emerging from the collision of a projectile of mass $M_P$ with a target of mass  $M_T\geq M_P$ can be described by a remnant body $M_r$ and continuous distribution of smaller fragments represented by a power law $\frac{dn}{dm} \sim n^{-b}$ up to the largest fragment size $M_F$. The mass of the remnant after the collision between the target and a projectile  can be described by \citep{morbidelli_asteroids_2009}

\begin{align}
\label{eq:M_R}
    M_R =\begin{dcases}
    [ -0.5 \times (\phi - 1) +0.5 ] \times (M_T +M_P) & \phi \leq 1 \\
        max\{ 0, [ -0.35 \times (\phi - 1) +0.5 ]\times (M_T +M_P)\} & \phi >  1\\
        \end{dcases}
\end{align}
where $\phi = Q/Q_d^*$ describes the ratio between the specific impact energy $Q = 1/2 \cdot v_{imp}^2 \cdot \mu$, where $\mu = \frac{M_T M_P}{M_T + M_P}$ is the reduced mass and $Q_d^*$ is the specific fragmentation energy of the target. The impact velocity $v_{imp}$ is given by the relative velocities between two swarms of planetesimal $(i,j)$ whose eccentricities and inclination follow Rayleigh distribution with mean $e_{i,j}$ and ${i_{i,j}}$: $v_{imp} = \sqrt{5/8 \Bar{e}^2+ 1/2 \Bar{i}^2}$ where $\Bar{e}^2 = e_i^2+e_j^2$ and $\Bar{i}^2 = i_i^2 +i_j^2$. 

The specific fragmentation energy describes the energy needed to fragment and disperse half of the target mass and can be described by \citep{benz_catastrophic_1999,benz_low_2000}
 \begin{align}
    Q_d^*(s) = Q_{0s} \Big(\frac{s}{cm}\Big)^{b_s} + Q_{0g}\rho_s\Big(\frac{s}{cm}\Big)^{b_g} + 9v_{esc}^2(s) , 
\end{align}
where $\rho_s$ is the bulk density of the planetesimals and the coefficients $Q_{0s}$, ${b_s}$, $Q_{0g}$ and $b_g$ can be found in \cite{kaufmann_influence_2023}. For simplicity, in this study, we utilise the fragmentation energy for icy bodies at $3$~km/s. In general, the specific fragmentation energy depends on the composition, relative impact speeds and whether we assume the target to be monolithic or a rubble pile. This can lead to a significant shift in the shape and magnitude of the specific fragmentation energy and therefore changes the evolution of the size distribution of the planetesimals (e.g. see \citealp{stewart_velocity-dependent_2009,leinhardt_collisions_2012,kobayashi_planetesimal_2018,krivov_debris_2018}), but due to the large uncertainties in the actual initial properties of the primordial planetesimals we did chose the description of \citep{benz_low_2000} used in many planet formation models including fragmentation \citep{sebastian_planetesimal_2019,lorek_growing_2022,kobayashi_planetesimal_2018}. For consistency, we calculate the specific fragmentation energy using the effective radius given by,
\begin{align}
    r_{eff} = \left(\frac{3(M_{T}+M_P)}{4\pi\rho_s}\right)^{1/3}.
\end{align}
For very high impact energies $\phi \gg 1$ $M_R$ can become negative and is set to $0$ and the target is considered to be pulverised and all the mass is lost.
The mass excavated from the target body is then given by $M_{ex} = M_{tot} - M_R$. Note that this description describes both accreting and disruptive collision as $M_R$ can be larger than $M_T$ for low-impact energies (i.e. for $Q/Q_d^* < 0.5$ for equal mass colliders). We distribute this excavated mass following a power law $dn/dm = m^{-p}$ between the largest fragment $M_F$ given by,

\begin{align}
    M_F = 8\times 10^{-3} \left( \frac{Q}{Q_d^*}e^{-(Q/4Q_d^*)^2} \right)\times M_{tot}
\end{align}
\noindent
and the minimum size considered which is two orders of magnitude lower than the smallest planetesimal size (i.e. corresponding to $r_{ext} = 10^{-2}$~cm) and the exponent being $p = 5/3$. The mass deposited in sizes below the smallest planetesimal bin is considered to be lost. For highly energetic collisions the mass of the largest fragment is set to $M_F = 0.5 M_R$.

The number of collisions between targets $i$ and projectiles $j$ during time $\delta t$ is given by \citep{ormel_understanding_2012},
\begin{align}
    N_{col}(i,j) = \delta t n_i n_j p_{coll}^{i,j} \sigma_{i,j},
\end{align}
\noindent
where $p_{coll}^{i,j}$ is the intrinsic collision probability \citep{morbidelli_asteroids_2009}, $\sigma_{i,j}$ the collisional cross section and $n_k$ the number of targets/projectiles respectively. In order to track the evolution of the size distribution we calculate the resulting change in mass for each bin according to the collisions with all the smaller projectiles only in their respective radial bin. To prevent spurious waves in the size distribution \citep{guilera_planetesimal_2014}, in each step, we reconstruct the size distribution below the minimal size via extrapolation down to $r_{ext} = 10^{-2}$cm and track the collisions of these particles with the larger ones as well. A comparison of our local fragmentation model presented here with the model described in \cite{guilera_planetesimal_2014} can be found in Appendix \ref{sec:app_frag}.

\subsection{Pebble accretion}
\label{sec:pebble_acc}

For our choice of planetesimal initial mass function, the largest bodies emerging from streaming instability are still too small to accrete pebbles efficiently if there is no migration trap for the pebbles to increase the encounter time \citep{lau_rapid_2022}. However, the mass of the most massive bodies emerging from SI can be quite close to the transition between the Bondi and Hill regimes, where pebble accretion becomes more efficient. Therefore we model the accretion of pebbles onto the embryo as described in \cite{liu_growth_2019,ormel_catching_2018} and \cite{liu_catching_2018}. For simplicity, we assume a radially constant flux of pebbles. We explore either a value that is constant in time of $F_{peb} =  50 M_\oplus/\text{Myr}$ or a time-dependent flux inferred from the evolution of the disk as described below. We assume pebbles have a fixed Stokes number of $\tau_s = 0.1$ in our nominal simulations as informed by dust growth simulation \citep{birnstiel_simple_2012,stammler_dustpy_2022}. We note that considering a fixed Stokes number is a good approximation when the pebbles growth is drift-limited, which usually happens in the outer part of the disk \citep[see for example][]{drazkowska_how_2021}. In addition, this approximation is often adopted in planet formation models that do not include detailed dust growth and evolution models \citep[e.g.][]{baumann_influence_2020}, which is the case in our study \citep{izidoro_formation_2021,liu_growth_2019,jang_growth_2022}.

The pebble accretion rate of the embryo can be described as a fraction of the incoming pebble flux as follows,
\begin{equation}
    \Dot{M}_{em} = \epsilon_{peb} F_{peb},
\end{equation}
where $\epsilon_{peb}$ is the pebble accretion efficiency which is described by 
\begin{equation}
    \epsilon_{peb} = \left(\epsilon_{2D}^{-2} + \epsilon_{3D}^{-2}\right)^{-0.5} \times f_{set} + \epsilon_{bal}\times (1-f_{set}),
\end{equation}
with $f_{set} = \text{exp}\left(-0.5\left(\frac{\Delta_v}{v_\star}\right)^2\right)$ being the settling factor determining the transition from the ballistic to the settling regime and $v_\star = \left(\frac{M_{em}}{M_\star \tau_s}\right)^{1/3}v_k$ is the corresponding transition velocity from the ballistic to the settling regime. The relative velocity between the planet and the pebble is given by
\begin{equation}
    \Delta_v = \left[ 1 + 5.7\left( \frac{q_p}{q_{hw/sh}}\right)\right]^{-1} v_{hw}+ v_{sh},
\end{equation}
where $v_{hw} = \eta v_k$ is the velocity contribution due to the particle drift and $v_{sh} = 0.52 (q_p\tau_s)^{1/3}v_k$ being caused by the Keplerian shear and factors $q_{hw/sh}=\eta^3/\tau_s$ and $q_p = M_{em}/M_\star$ govern the transition between the two regimes. The accretion in the settling regime in the 2D and 3D regime is then given by
\begin{equation}
    \label{eq:acc_2d}
    \epsilon_{2d} = 0.32\sqrt{\frac{q_p \Delta_v}{\tau_s\eta^2 v_k}}f_{set},
\end{equation}
and 
\begin{equation}
    \epsilon_{3d} = 0.39 \frac{M_{em}}{\eta h_{peb} M_\star} f_{set}^2
\end{equation}
respectively, and $\epsilon_{bal} = \frac{R_p}{2\pi\tau_s\eta r_p}\sqrt{\frac{2 q_p r_p}{R_p}+\left(\frac{\Delta_v}{v_k}\right)^2}$ is the ballistic cross-section. The pebble aspect ratio is given by $h_{peb} = \sqrt{\frac{\alpha_z}{\alpha_z + \tau_s}}h_{gas}$ where $h_{gas}$ is the aspect ratio of the gas disk. We use a value $\alpha_z = 10^{-4}$ to describe the vertical mid-plane turbulence which is different from the turbulent viscosity $\alpha$ used to evolve the gas disk \citep{pinilla_growing_2021}. 

\begin{figure}
    \centering
    \includegraphics[width=\hsize]{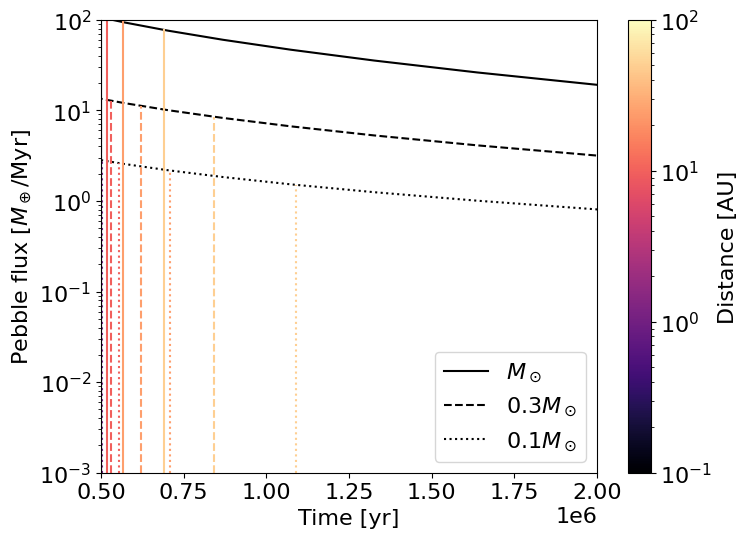}
    \caption{The resulting pebble fluxes from Eq. \eqref{eq:timed_fpeb} for $Z_0 =0.01$ around a solar mass star (\textit{solid}), a $0.3M_\oplus$ star (\textit{dashed}) and a $0.1M_\oplus$ star (\textit{dotted}) along with the $t_0$ as described by Eq. \eqref{eq:t_0final} for filaments at different separations form the star}
    \label{fig:flux_peb_t}
\end{figure}

Additionally, instead of assuming a pebble flux that is constant in time, we can account for the evolution of the pebble flux described in \cite{lambrechts_forming_2014}. Therefore, we define the growth radius $r_g$ which is the distance from the star where the dust has grown to sizes where its growth is halted by its inward drift i.e. where its growth timescale is equal to its drift timescale,
\begin{equation}
    t = \text{ln}(R_{drift}/R_0)t_{growth} \approx \xi\frac{4}{\sqrt{3}\epsilon_p Z_0 \Omega_k},
\end{equation}
which results in a growth radius of
\begin{equation}
    r_g(t) = \left(\frac{3}{16}\right)^{1/3} (GM_\star)^{1/3}(\epsilon_dZ_0)^{2/3}t^{2/3}.
\end{equation}
As this growth radius moves outwards we calculate the resulting mass flux from the inward drifting pebbles which we assume to be radially constant inside the growth radius and is described by
\begin{equation}
    \label{eq:timed_fpeb}
    F_{peb} = 2\pi r_g \frac{\text{d}r_g}{\text{d}t} \Sigma_{d,0}(r_g),
\end{equation}
where $\Sigma_{d,0}$ is the initial dust surface density at $t=t_0$ which is linked to the initial gas surface density via the disk metallicity via  $\Sigma_{d,0} = Z_0*\Sigma_{g,0}$. The resulting pebble flux for our chosen disk model around a solar a $0.3$ and a $0.1  M_\odot$ mass star with $Z_0=0.01$ can be seen in Fig. \ref{fig:flux_peb_t}. Note that due to the time offset $t_0$ in our initial conditions that accounts for the time for the filament to form, the pebble flux has to account for the same time offset to be consistent. This is described in further detail in Section \ref{sec:t0}. Since the characteristic planetesimal mass in our setup accretes pebbles well in the ballistic regime which leads to negligible accretion rates, we only consider the accretion onto the embryo.

A good heuristic to identify which bodies grow significantly from pebble accretion is to compare the mass of the accreting body with the following characteristic masses. The onset mass $M_{on}$, marks the transition from the pebbles being accreted in the ballistic to the Bondi/Hill regime and it can be derived by equating the encounter time of the pebbles with the embryo and their friction time. Below this mass, pebble accretion is slow as the cross-section of the accreting body is not enhanced by gas drag. This transition happens at a mass of:
\begin{equation}
    M_{on} = \frac{1}{4} \tau_s \eta^3 M_\star.
\end{equation}

Additionally, we introduce the transition mass $M_{tr}$. It describes the mass where the accretion regime of the pebbles changes from the Bondi to the Hill regime \citep{ormel_catching_2018} following the description in \cite{pessah_emerging_2017} it is described by:
\begin{equation}
    \label{eq:M_tr}
    M_{tr} = \frac{\eta^3 M_\star}{\sqrt{3}}.
\end{equation}
Above the transition mass $M_{tr}$ the accretion of pebbles is enhanced (in the 2D regime) as can be seen in Eq. \eqref{eq:acc_2d}.

\begin{figure}
    \centering
    \includegraphics[width=\hsize]{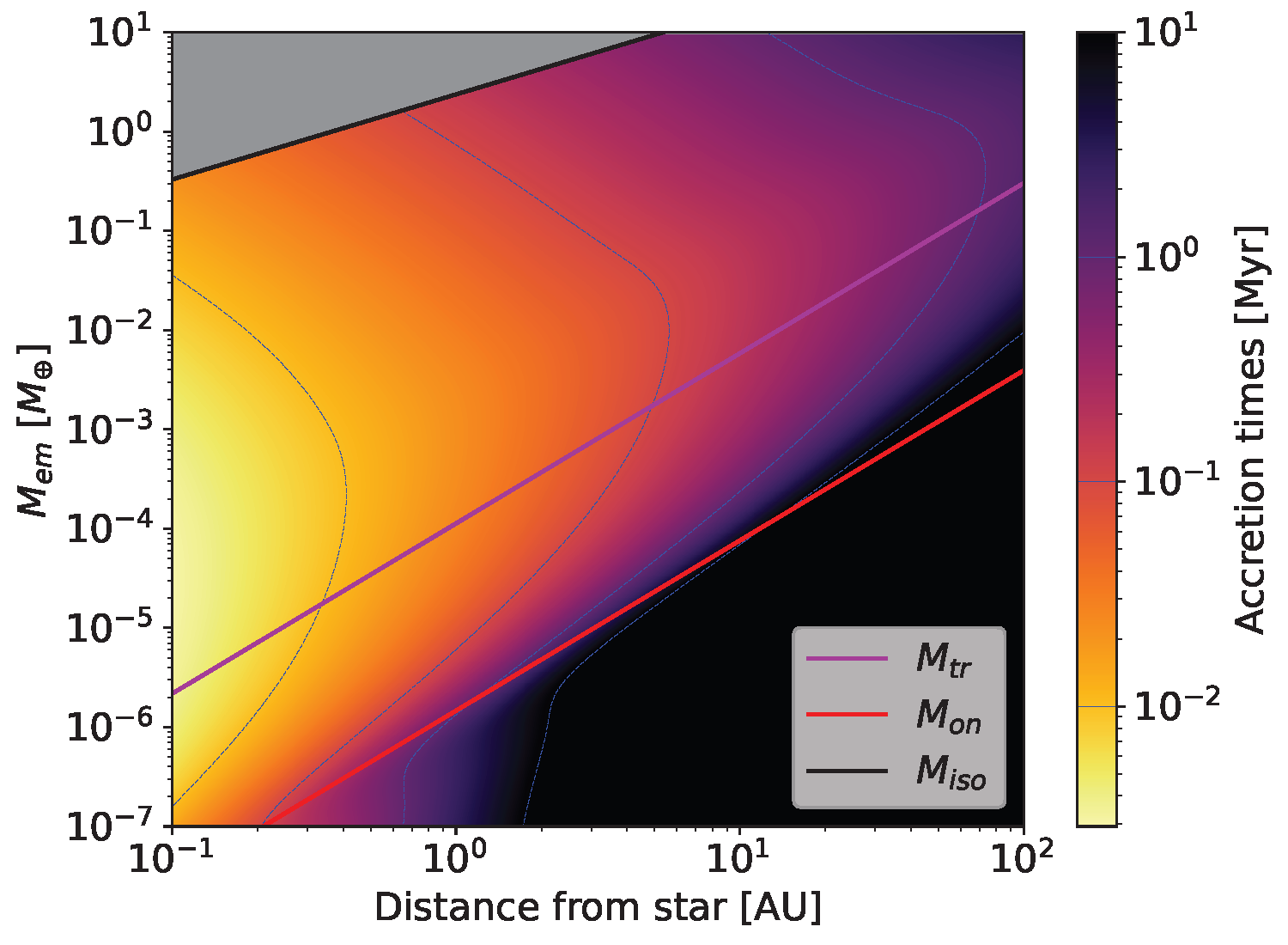}
    \caption{The accretion timescale ($M_{pl}/\Dot{M}_{pl}$ [Myr]) for a fixed pebble flux and $\tau_{peb}=0.1$ as a function of the embryo mass $M_{em}$ and distance from the star with the different transition regimes.}
    \label{fig:pebble_acc_eff}
\end{figure}

Finally, when the embryo is large enough and reaches its isolation mass $M_{iso}$ it carves a gap in the gas disk thus reversing the local pressure gradient of the gas and hereby cuts itself off from the incoming pebble flux which happens at \citep{bitsch_pebble-isolation_2018,ataiee_how_2018}, 
\begin{equation}
    M_{iso} = 25M_\oplus \left(\frac{H/r}{0.05}\right)^3 \left(0.34\left(\frac{-3}{\text{log}_{10}\alpha_t}\right)^4 +0.66\right)\left( 1 - \frac{\frac{d\text{ln}P}{d\text{ln}r} + 2.5}{6}\right),
\end{equation}
where $\frac{d\text{ln}P}{d\text{ln}r}$ is the unperturbed local logarithmic pressure gradient. To illustrate the different accretion regimes and the corresponding pebble accretion times $M/\dot{M}_{peb}$ in the mass ranges considered for our initial conditions, we plot the pebble growth times for a pebble flux of $50~\text{M}_\oplus/\text{Myr}$ and $\tau_s = 0.1$ for our chosen disk model which can be seen in Fig. \ref{fig:pebble_acc_eff}. We can clearly see both transitions from the ballistic to the Bondi regime and from the Bondi to the Hill regime from the change in the accretion time dependency on the embryos's mass. Here we can clearly see the efficient accretion in the Hill regime where the accretion time only weakly depends on the embryo's mass.

\subsection{Initial conditions and disk model}
In this paper, we follow the growth after a localised formation of planetesimals as has been investigated in multiple previous works \citep[e.g.][]{lorek_growing_2022,liu_growth_2019}. We do not simulate the concentration of the dust and subsequent collapse to form the initial ring of planetesimals and assume that all the planetesimals are formed at time $t_0$ in our simulation. However, to account for the fact that these processes have to take place before the start of our simulation we chose the remaining initial conditions to reflect this which we discuss in further detail in Section \ref{sec:t0} Note that this means we have different $t_0$ for the different locations in the disk.

\subsection{Gas disk}
\label{sec:initial}

To describe the initial gas disk and its evolution we follow the viscously evolving $\alpha$ disk \citep{lynden-bell_evolution_1974} using the self-similar solution. The temperature profile of the disk is chosen to be \citep{ida_radial_2016}:
\begin{align}
     T = 150K \: (M_\star/M_\odot)^{-1/7} (L_\star/L_\odot)^{2/7} (r/\text{AU})^{-3/7}.
\end{align}
We chose a constant $\alpha = 10^{-2}$ throughout the disk. This results in a power law dependence of the viscosity of $\nu = \alpha c_s^2\Omega_K^{-1}$. The evolution of the surface density for our chosen temperature profile and $\nu \sim r^{\gamma}$ ($\gamma = 15/14$) can be described by the self-similar solution which is given by: 
\begin{align}
    \Sigma_{g} = \frac{\Dot{M}_{acc,0}}{3\pi \nu(r_1)} (r/r_1)^{-\gamma} \tau^{-(5/2-\gamma)/(2-\gamma)} \exp \left[-\frac{1}{\tau}\Big(\frac{r}{r_1}\Big)^{2-\gamma}\right]
\end{align}
The characteristic evolution timescale is given by $\tau = t/(r_1^2/(3(2-\gamma)\nu(r_1))) +1$ and $r_1$ is the characteristic radius \citep{armitage_protoplanetary_2019}. The initial mass accretion rate is chosen to be $M_{\star,0} = 10^{-7} M_\odot/yr$ at $0.5$ Myr. The total disk mass is set to be $M_{gas} = 0.1 M_\star$ at time $t=0$ which lets us calculate the characteristic radius $r_1$ by integrating the surface density profile (from 0.1 AU to infinity) and choosing $r_1$ so the total gas mass matches the chosen value. This yields a characteristic radius of $\sim 72$ AU for a solar mass star.

To model the formation around stars with different masses we have to adapt the disk model and initial conditions. As lower mass stars show lower accretion rates we scale the gas disk model with the stellar mass according to \cite{hartmann_accretion_2016} 
 linearly $\Dot{M} \propto M_\star$. We keep the disk to stellar mass ratio constant for all stellar masses this means the characteristic radius $r_1$ is different for different stellar masses. We scale the luminosity as $L_\star \propto M_\star^{1.5}$ which is consistent with the scaling derived for young stars $<10$ Myr that show a range of $1-2$ or the exponent in the $L_\star$--$M_\star$ relation for these young stars \citep{liu_pebble-driven_2020}.

\subsection{Filament}
\label{sec:fil}
\begin{figure}
    \centering
    \includegraphics[width=\hsize]{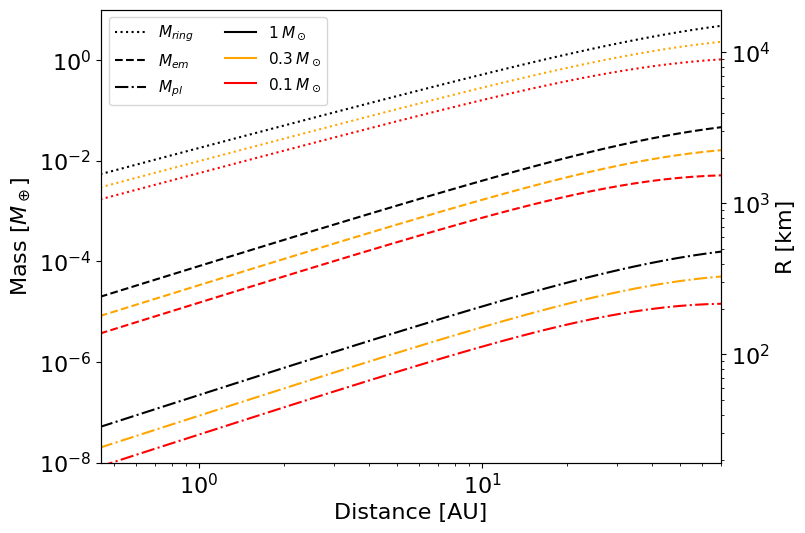}
    \caption{The total ring mass (\textit{dotted}), the mass of the single largest (\textit{dashed}) and characteristic (\textit{dashdotted}) planetesimal mass for different stellar masses (\textit{colour}) for filaments with different separations from the star}
    \label{fig:mass_init}
\end{figure}

The localized ring of planetesimals modelled in this work is assumed to have formed by streaming instability \citep{liu_growth_2019,gerbig_requirements_2020}. The typical width of a dense filament of pebbles and the resulting planetesimals forming in them is typically determined by the pressure gradient of the disk $\eta = -0.5 h^2 \frac{d\text{ln}P}{d\text{ln}r}$ and the separation from the star $r_0$ as $\Delta w = \eta \cdot r_0$. By postulating a solid-to-gas ratio in the filament $Z$ and a planetesimal formation efficiency factor $p_{eff}$ we can calculate the total mass of planetesimals created in such a filament. The total mass of planetesimals is given by $M_{fil} = 2\pi r_0 \Delta w \Sigma_{gas}Zp_{eff}$. The Initial mass function (IMF) of the planetesimals created in these filaments can be described by the cumulative number distribution given by \citep{schafer_initial_2017}
\begin{align}
    \frac{N_\geq(m)}{N_{tot}} = \left(\frac{m}{m_{min}}\right)^{-p} \text{exp} \left[\left(\frac{m_{min}}{m_p}\right)^q - \left(\frac{m}{m_p}\right)^q\right],
\end{align}
where the characteristic planetesimal mass $m_p$ is given by,
\begin{align}
    \label{eq:m_pl}
    m_p = 5\times 10^{-5} M_\oplus \left(\frac{Z}{0.02}\right)^{1/2} \left(\gamma_s \times\pi\right)^{3/2} \left( \frac{h}{0.05}\right)^{3} \left(\frac{M_\star}{M_\odot}\right)
 \end{align}
and $\gamma_s = 4\pi G \rho \Omega_k^{-2}$ is a self gravity parameter of the gas \citep{liu_tale_2020}. For the parameters chosen ($q = 0.4$ and $p = 0.6$), the mass budget of the planetesimals is dominated by bodies of the characteristic mass $m_p$ whereas the number of bodies is dominated by the smallest bodies $N_{tot} = M_{fil}/m_{min}$ which is chosen to be $m_{min} = 10^{-3} m_p$ in our setup \citep{lorek_growing_2022}. This allows us to calculate the largest single body generated in such a filament which is given by $N_\geq(M_{em}) = 1$. We use this body as the planetary embryo i.e. the initial seed for our forming planet's core placing it at $r_0$ to track its subsequent growth. As the IMF is strongly peaked around its characteristic mass $m_p$ we initialize the planetesimals as a single-sized population of planetesimals of size $r_{init} =  \sqrt[3]{3 m_p /(4\pi \rho)}$. The initial surface density of the planetesimals given by

\begin{equation}
    \Sigma_p = \begin{dcases}
    M_{fil}/(2\pi r_0^2\eta), & \text{if } |r-r_0| < \eta \times r_0/2 \\
    0, &\text{otherwise.}\\
    \end{dcases}
\end{equation}
The initial eccentricities and inclinations of the planetesimals are given by $e = 2i = \eta/2$ \citep{lorek_growing_2022}. Although they are not well constrained from the formation, we know that they have to be smaller than the width of the filament. To visualise the mass scales of the different components described above we plot the characteristic mass of the planetesimal along with the initial embryo mass and the mass of the entire ring in Fig. \ref{fig:mass_init} for the different stellar masses for our nominal choice of filament metallicity of $Z=0.1$.

The initial masses of the planetesimals, the embryo and the ring increase monotonically with increasing semi-major axis and are lower for lower stellar masses. The ratio of the largest body to the characteristic planetesimal mass stays roughly constant throughout the disk with a ratio of $\approx 330$. Note that the initial mass of the embryo is significantly lower than the initial embryo mass used in most planet formation models (typically $0.01 M_\oplus$).

\subsection{Initial time}
\label{sec:t0}
As we assume the planetesimals to be formed at the start of our simulations we have to account for the time that has passed for the dust to grow into pebble sizes and the time required for the pebbles to collapse into planetesimals. We assume the dust growth to commence in a class-II disk with a typical age of $t_{classII} = 0.5$ Myr \citep{williams_protoplanetary_2011}. To estimate the growth timescale we follow the approach of \cite{lorek_growing_2022}. The growth timescale of the dust \citep{birnstiel_simple_2012} can be described by
\begin{equation}
    t_{gr} = \frac{2}{\sqrt{\pi}\epsilon_p Z_0 \Omega_K},
\end{equation}
where the sticking efficiency given by $\epsilon_p = 0.5$ and the $Z_0$ is the dust to gas ratio in the disk. Then the time required for the dust of initial size $r_0$ to grow to $r_{max}$ is simply,
\begin{align}
    t_{peb} = t_{gr} \text{log}\left(r_{max}/t_0\right),
\end{align}
the growth of the pebbles is limited by two processes: drift and fragmentation. The upper size limit of the pebbles in terms of their Stokes number is described by \citep{birnstiel_simple_2012},
\begin{align}
    \label{eq:st_peb_frag_drift}
    \tau_{s} = \begin{dcases}
        \frac{3\sqrt{\pi}\epsilon_p Z}{4\eta} &\text{drift limit}\\
        \frac{v_{frag}^2}{2\alpha_tc_s^2} &\text{fragmentation limit}\\
    \end{dcases}
\end{align}
where $v_{frag}$ is the fragmentation velocity which we consider to be $1$  m/s in this work. The physical size of the pebbles can then be calculated from their Stokes number and assuming Epstein drag via
\begin{equation}
    r_p = \frac{2\tau_s\Sigma_g}{\pi\rho_s}.
\end{equation}
The time that it takes for the pebbles to collapse into the final planetesimals is on the order of tens to several thousand orbital periods \citep{yang_concentrating_2017,li_demographics_2019}. Therefore we use $t_{SI} = 500* 2\pi/\Omega_K$. This means the initial time of our simulations is considered to be,
\begin{align}
    \label{eq:t_0final}
    t_0 = t_{classII} + t_{peb} + t_{SI}.
\end{align}
For our nominal setup around a solar-type star this results in a time offset from $0.5$ Myr in the inner disk up to $\sim 1\text{Myr}$ at $100 \text{AU}$. The results shown in the following section all display the time relative to the initial time $t_0$ of the simulation.

\section{Results}
\label{sec:Results}
\begin{figure}
    \centering
    \includegraphics[width=\hsize]{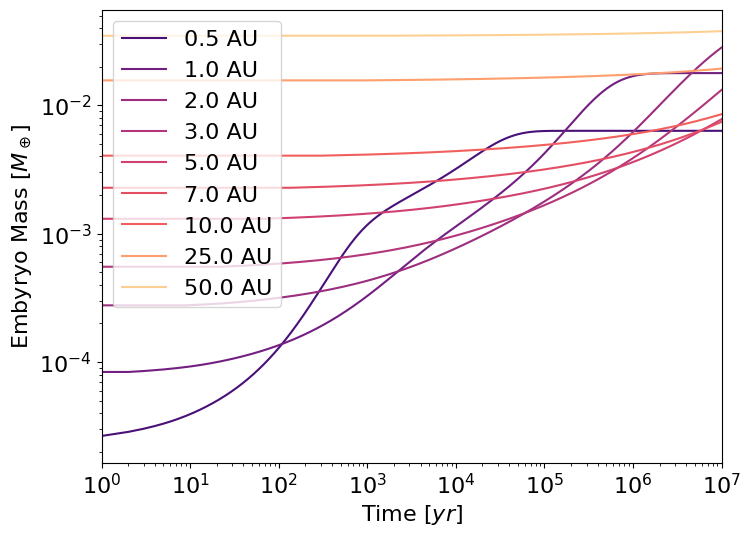}
    \centering
    \includegraphics[width=\hsize]{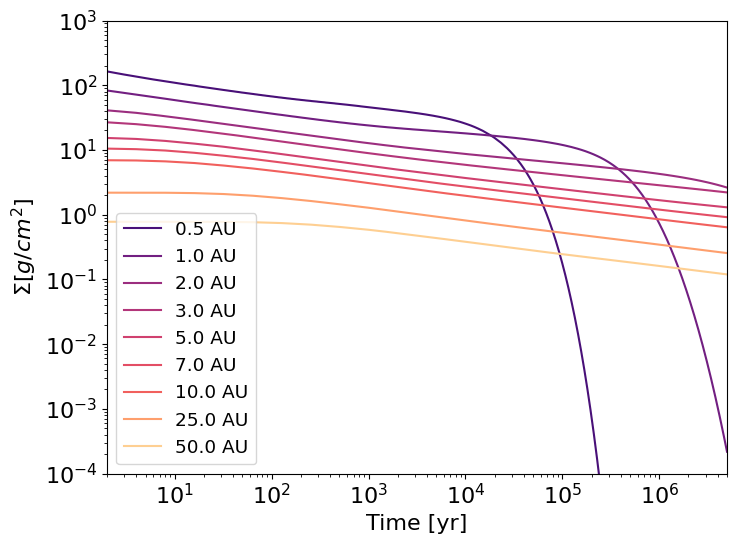}
    \caption{The embryo masses for filaments at different separations from the star over time (\textit{top}) and the surface density of planetesimals in the ring (\textit{bottom}) as a function of time}
    \label{fig:nominal_track}
\end{figure}

 We probed the formation of embryos in filaments at different locations ranging from $0.5$ to $50$ AU from their initial time $t_0$ to $10$ Myr which serves as a reasonable upper bound for the gas disk lifetime. The growth of the embryos and the evolution of the surface density of the planetesimals at the different separations from the star for our nominal model can be seen in Fig. \ref{fig:nominal_track}. The parameters of our nominal setup are a metallicity of $Z=0.1$ for the filament with a $M_{gas} = 0.1M_\odot$ gas disk around a solar mass star, and we do without the inclusion of pebble accretion and fragmentation. The remaining parameters of the nominal simulation can be found in the Appendix in Table \ref{tab:nominal_params}. 
 
\begin{figure}
    \centering
    \includegraphics[width=\hsize]{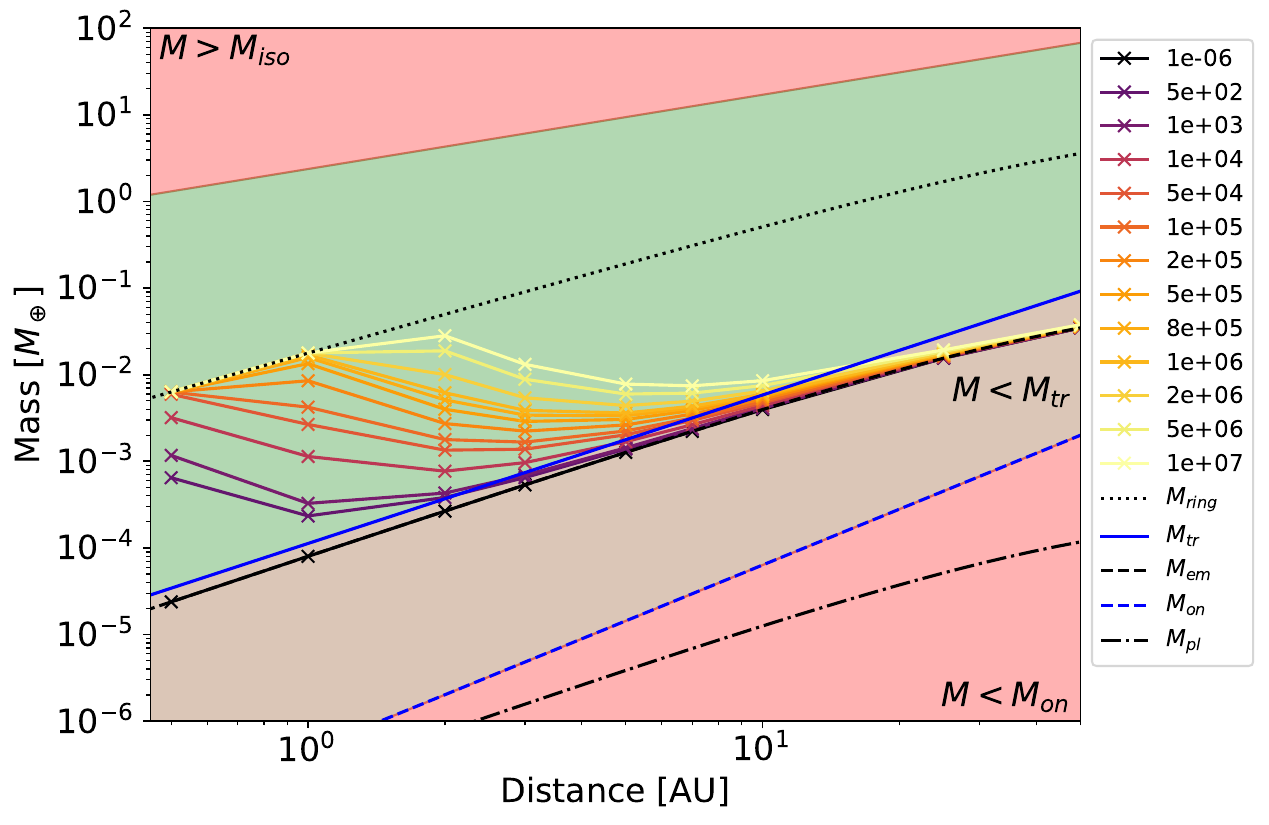}

    \caption{The mass of the largest body in the ring (embryo) (\textit{solid}), the initial characteristic planetesimal mass (\textit{dash doted}) and ring mass (\textit{dotted}) where the background colours refer to the different pebble accretion regimes (\textit{red}: ballistic/isolation regime, \textit{brown}: Bondi regime and \textit{green}: Hill regime)}
    \label{fig:M_sma_nom}
\end{figure}

The initial mass of the embryos is ascending with increasing separation from the star as dictated by our initial conditions (see Fig. \ref{fig:mass_init}), but due to the higher surface density of planetesimals and higher orbital frequency closer to the star, the embryos closer in can grow significantly faster and accrete all the mass in their filament, whereas with increasing distance the growth of the largest body is significantly slower, making it almost negligible at $50$ AU. The surface density of the planetesimals in the ring starts to decrease even at early times due to the diffusion of the planetesimals which is caused by their mutual gravitational interaction, which acts as a further barrier to the growth of the embryo as it directly relates to the accretion rate which is shown in the bottom panel of Fig. \ref{fig:nominal_track}. In order to get a better idea of how the filaments at different separations evolve in time, we plot the embryo masses throughout time for the filaments at the different semi-major axes in Fig. \ref{fig:M_sma_nom}. We also plot the characteristic mass of the planetesimals along with the total mass contained in the filament. To better visualize where pebbles could be accreted efficiently (although pebble accretion is not considered for this set of simulations) we plot the different efficiency regimes according to the transitions from ballistic to Bondi and Bondi to Hill respectively as described in Section \ref{sec:pebble_acc}.
As we can see the initial mass of the largest body i.e. the embryo is quite close to the transition mass (by a factor of $\approx 1.5$) for the filaments up to $10$ AU. Even so, the growth of the largest body from planetesimal accretion is only significant enough to reach the transition mass up to $\sim 10$ AU. Farther out there is virtually no growth and the embryo fails to accrete any significant amount of mass from the planetesimal ring. However, in the inner disk, the embryo is able to accrete the entire mass of the filament up to $\sim 2$ AU. Whereas, in the intermediate regime, the growth is too slow to accrete all of the material. This growth pattern is consistent with previous studies \citep{lorek_growing_2022} that also show this stark radial dependence of embryo growth.

To investigate the impact different physical processes have on the early growth in these filaments, we first run simulations with the same setup as above, but including additional physical effects, for example, fragmentation and the effect diffusion has on early growth. Additionally, we investigate the formation around lower-mass stars when considering planetesimal accretion only. Then, we will investigate how the early growth is impacted when we consider the accretion of pebbles concurrently. Finally, we will discuss the implications of our results for the timing of core formation and how it can inform the initial conditions of global formation models.

\subsection{Fragmentation}
\begin{figure}
    \centering
    \includegraphics[width=\hsize]{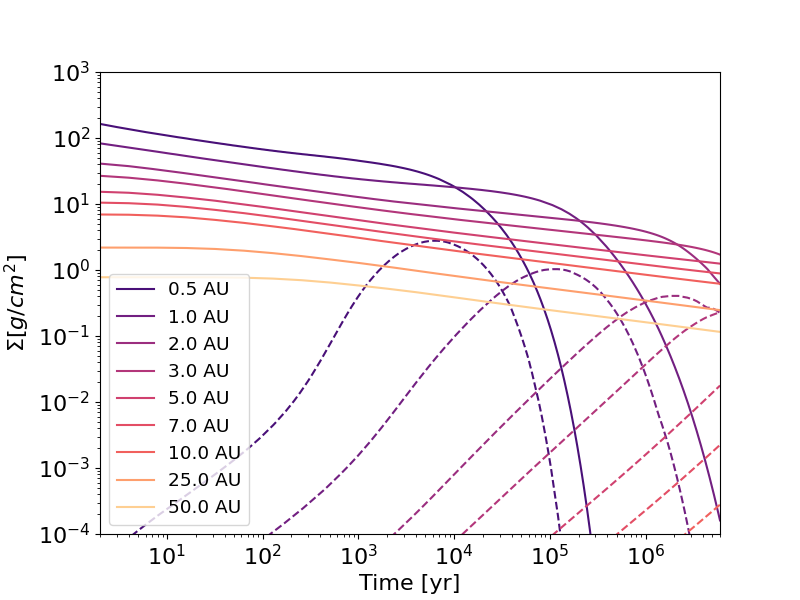}
    \caption{The surface density of the planetesimals (\textit{solid}) and fragments (\textit{dashed}) for the filaments at different separations}
    \label{fig:sigma_pop_frag}
\end{figure}

As described by \cite{lorek_growing_2022} the consideration of fragmentation can have an effect on the mass growth and final mass of the embryo forming in these rings. To investigate this we also run a set of simulations that include the effect of fragmentation with the collision model as described in section \ref{sec:frag_theory}. To illustrate the effect fragmentation has on the growth of embryos in Fig. \ref{fig:sigma_pop_frag} we show the surface densities of initial planetesimals and the fragments i.e. the sum over all the planetesimal mass bins smaller than the initial planetesimal mass that get created via collisions in the ring in Fig. \ref{fig:sigma_pop_frag}.

\begin{figure}
    \centering
    \includegraphics[width=\hsize]{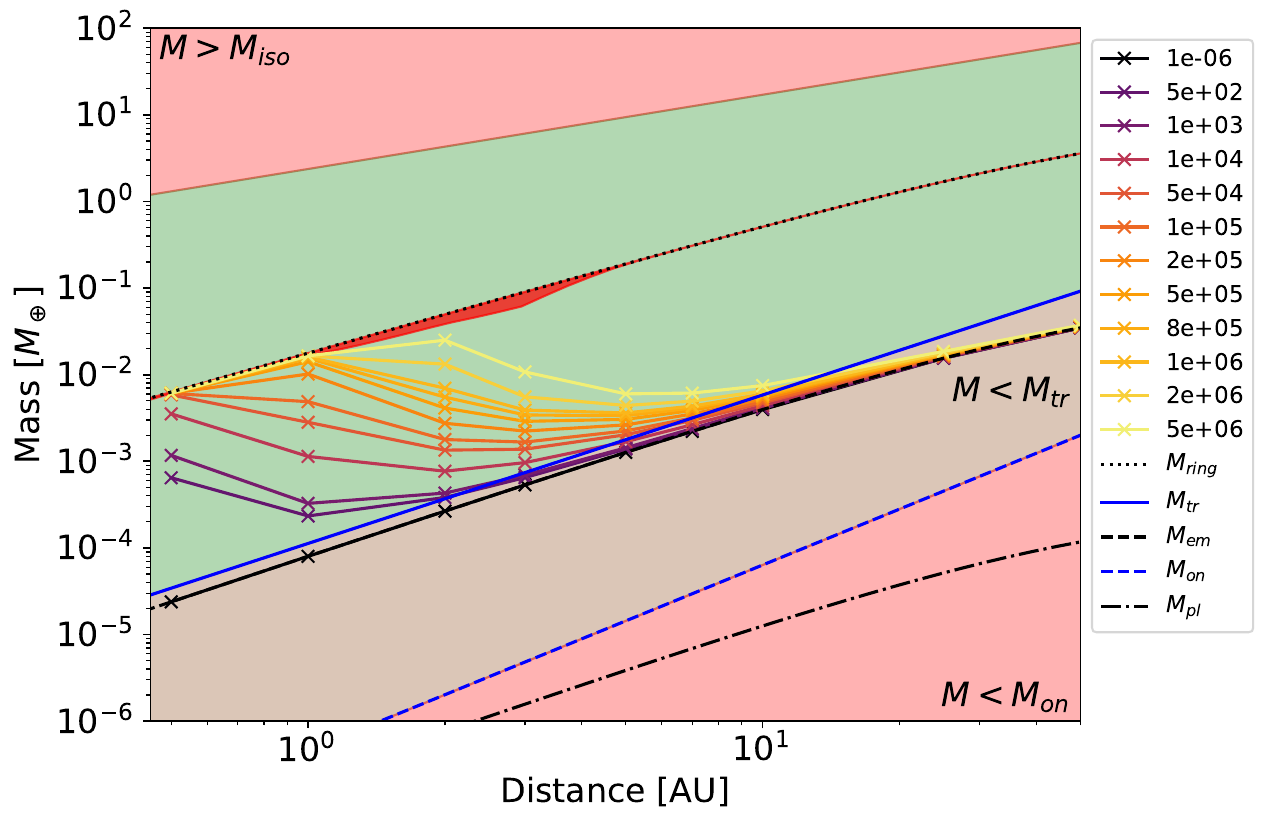}
    \caption{Same as Fig. \ref{fig:M_sma_nom} with the addition of the mass lost from the collisions among planetesimals (\textit{dark red})}
    \label{fig:M_sma_frag}
\end{figure}

As we can clearly see both the amount of fragments and the speed at which they are created is significantly shorter in the inner disk whereas virtually no fragments are created in the outer disk. This can be explained by the longer collision timescales due to the lower surface densities, scaling with separation and bigger initial planetesimal size leading to a higher specific material strength of the initial planetesimals \citep{benz_catastrophic_1999}. The resulting growth of the filaments including fragmentation can be seen in Fig. \ref{fig:M_sma_frag}. The final mass of the largest body within $1$ AU is smaller than its non-fragmenting counterpart due to the fact that some of the mass excavated in the collisions is deposited into dust (i.e. mass deposited in sizes below $r_{min}$) which is considered to be lost in our model for the purposes of accretion. However, the growth of the filaments at intermediate separations between $\sim 1-5$ AU is enhanced when the embryo approaches the filament mass as the embryo starts to stir up the relative velocities of the planetesimals, this leads to higher final masses for the bodies as the fragments are easier to accrete as they are more tightly bound to the gas. As indicated by the number of fragments produced and the growth track fragmentation virtually plays no role in the growth of the embryo in the outer disk which is in line with previous findings \citep{kaufmann_influence_2023,lorek_growing_2022}. Note that due to computational constraints, we have evolved these simulations only for $6$ Myr.

\subsection{Effect of diffusion}
\begin{figure}
     \centering
     \includegraphics[width=\hsize]{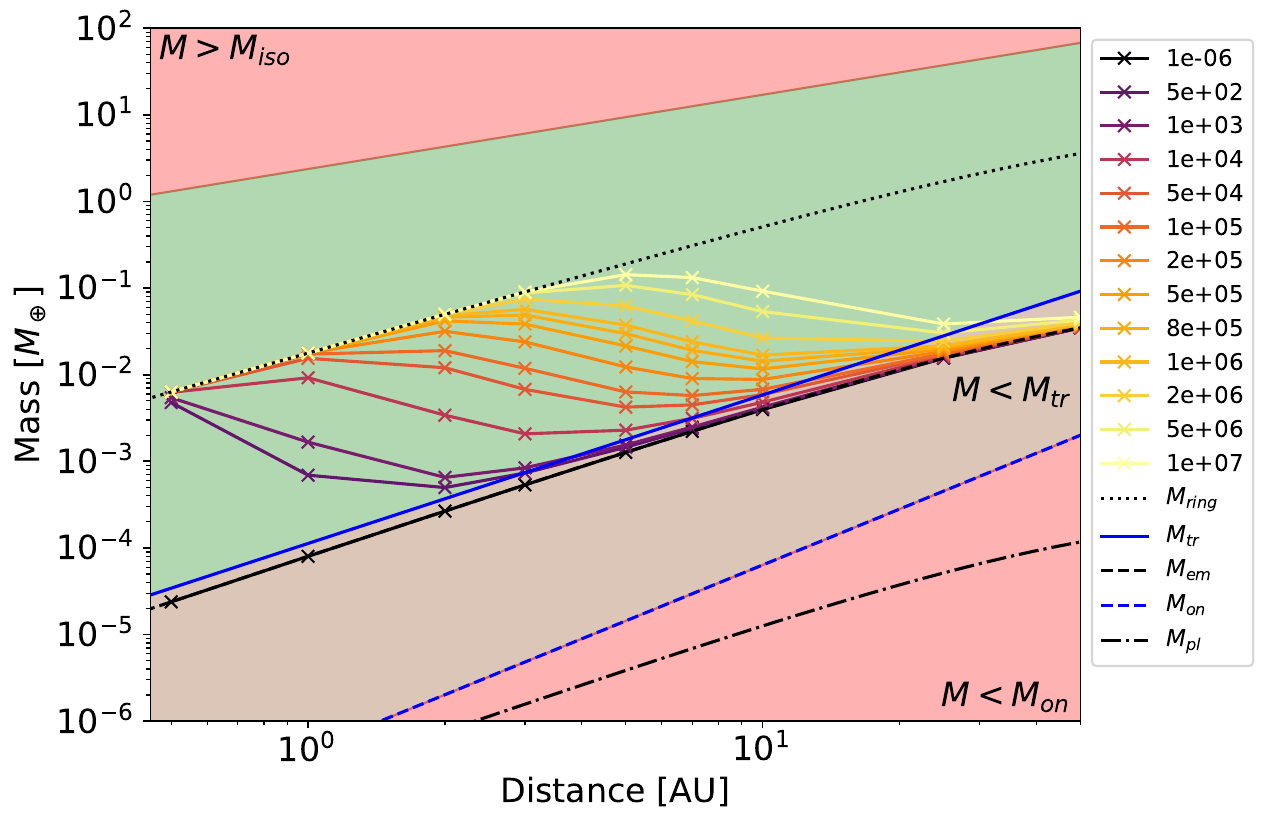}
     \caption{Same as Fig. \ref{fig:M_sma_nom} but excluding the diffusive widening of the planetesimal ring}
     \label{fig:m_sma_nodiff}
 \end{figure}

The diffusive widening of the ring of planetesimals reduces their surface density as can be seen in Fig \ref{fig:nominal_track} which slows down the growth of the embryo. To understand the importance of this effect we performed a set of simulations where we set the diffusion term in Eq. \eqref{eq:cont_solids} to zero. Although nonphysical, it illustrates a setup closer to the formation in structured disks that act as migration traps due to a change in the gas pressure gradient \citep{jiang_efficient_2022}. The growth of the different filaments without diffusion can be seen in Fig. \ref{fig:m_sma_nodiff}.

 The growth of the embryos is significantly enhanced when the diffusion is not considered, allowing the largest body to accrete the entire mass contained in the filament up to $\sim 7$ AU. Additionally, it also allows the largest body to grow to the transition mass for larger separation up to $25$ AU. This clearly demonstrates a difference in growth patterns when we consider a radially localized distribution of planetesimals.

\subsection{Varying stellar mass}
\begin{figure}
    \centering
    \includegraphics[width=\hsize]{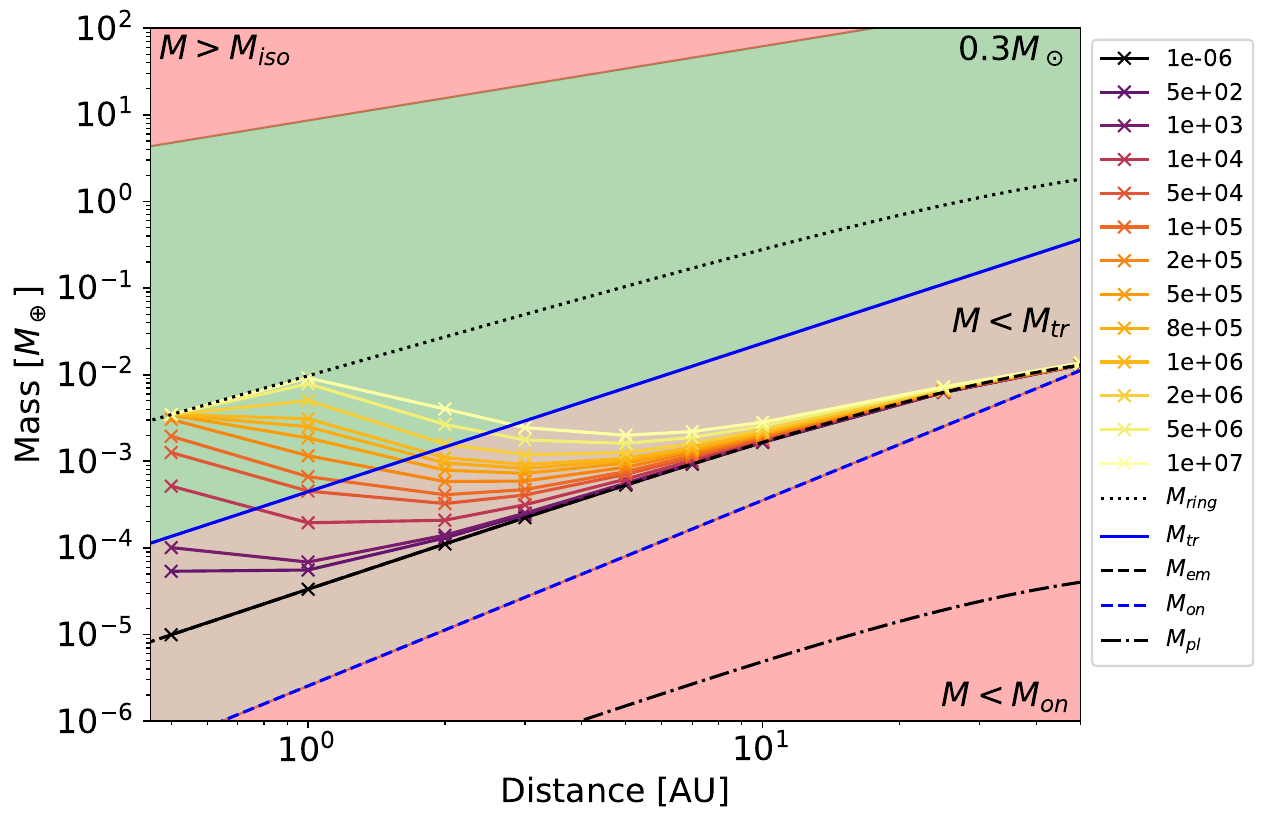}
    \centering
    \includegraphics[width=\hsize]{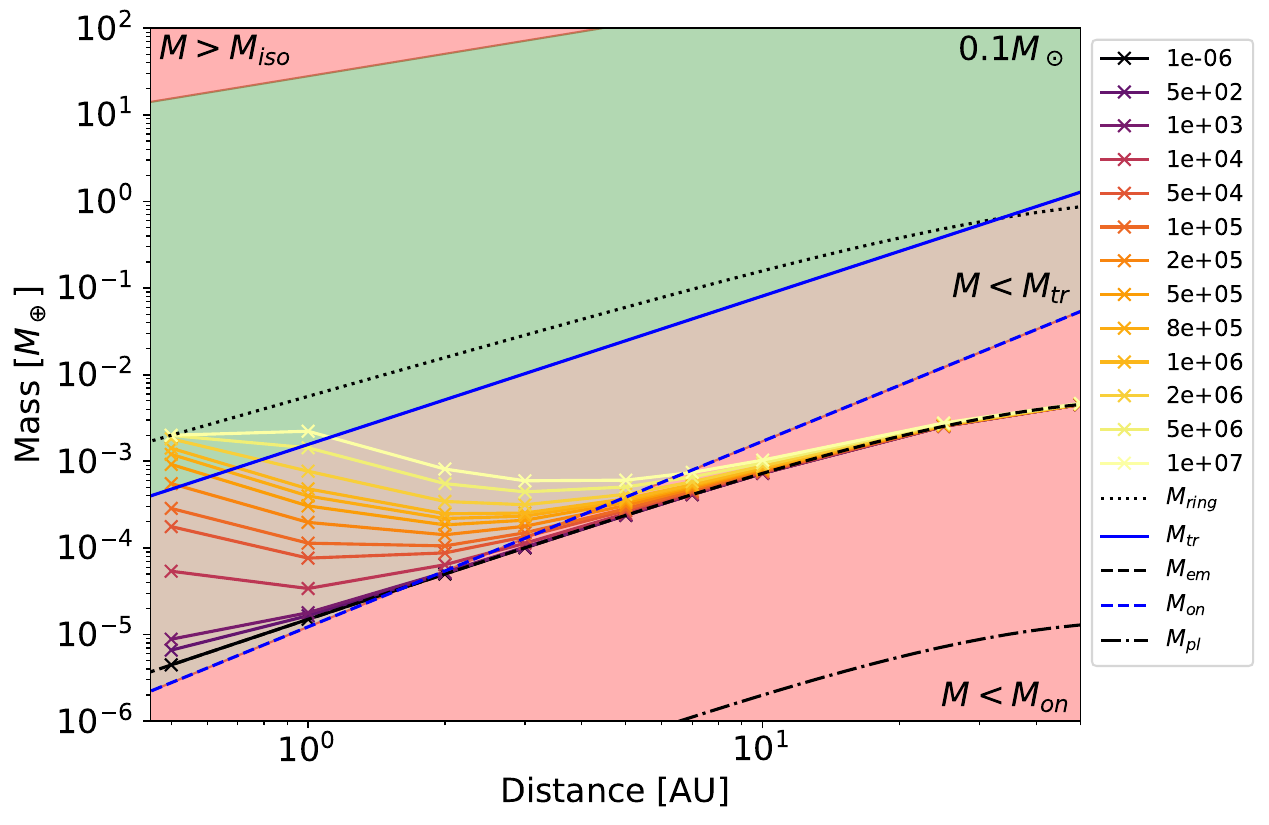}
    \caption{Same as Fig. \ref{fig:M_sma_nom} but for filaments around a $0.3 M_\odot$ mass star (\textit{top}) and around a $0.1 M_\odot$ mass star (\textit{bottom})}
    \label{fig:M_sma_0.3}
\end{figure}

As planet formation strongly depends on the mass of its host star, we additionally run a set of simulations for a host star with a mass of $0.3$ and a $0.1 M_\odot$. Where the changes in the disk model and initial conditions have been outlined in Section \ref{sec:initial} and the change in mass for the ring and planetesimals can be seen in Fig. \ref{fig:mass_init}. Note that in order to keep the disk star mass ratio constant (i.e. $M_{\text{gas}} = 0.1 M_\star$) the characteristic radius of the gas disk varies for different stellar masses. In Fig. \ref{fig:M_sma_0.3} we show the growth of the filaments for a $0.3 M_\oplus$ (\textit{top}) and a $0.1 M_\odot$ (\textit{bottom}) mass star at different separations, using the nominal setup, can be seen in Fig. \ref{fig:M_sma_0.3}. 

As we can clearly see the growth around both the $0.3$ and $0.1$ $M_\odot$ mass star are significantly reduced. For the $0.3 M_\odot$ star the filaments up 3 AU are able to reach the transition mass whereas for the $0.1 M_\odot$ star, this only happens up to $\sim 1$AU. This can easily be explained due to how the chosen initial conditions and the characteristic masses scale with the mass of the host star, i.e. the embryos need to grow significantly more in order to reach the transition mass $M_{tr}$ for lower stellar masses. Although the ring properties are dependent on the distance from the star the following scaling relations we derive are valid for planetesimal rings with separations up to $10$ AU from the star. Here we will outline how the different components scale with the mass of the host star to explain the aforementioned differences in early growth of the embryo for varying stellar mass. Further out there is a radial dependence of these scaling laws due to the different characteristic radii and growth times (leading to different $t_0$) so the following relations are only applicable to the inner disk where the initial gas surface density scales as $\Sigma_g \propto M_\star^{1.215}$ (which was fitted to the initial disk profiles around different stellar masses). As a result of this the characteristic planetesimal mass given in Eq. \eqref{eq:m_pl} scales as $M_{pl} \propto M_\star^{0.78}$ with the stellar mass of the host star. This leads to a scaling of the mass of the largest body according to $M_{em} \propto M_\star^{0.72}$ (Note that these relations are not derived fully analytically as the initial planetesimal surface density and the embryo mass are calculated from implicit equations). For our chosen disk model the transition mass scales as $M_{tr} \propto M_\star^{-8/7}$. The onset mass for pebble accretion scales either as $M_{on}\propto M_\star^{-8/7}$ or $M_{on}\propto M_\star^{-10/7}$ depending if we choose to consider the Stokes number for the pebbles to be fixed or if it's given by the fragmentation size limit in Eq. \eqref{eq:st_peb_frag_drift}. This makes it quite apparent that the formation of massive embryos around higher-mass stars is a lot easier as the initial masses of the embryos are higher and transitions of the pebble accretion regimes happen at lower masses. This illustrates well that the largest planetesimals are smaller around lower-mass stars and the characteristic masses of pebble accretion are larger making it harder to form massive embryos around these stars. 

\subsection{Including pebble accretion}
\begin{figure}
    \centering
        \includegraphics[width=\hsize]{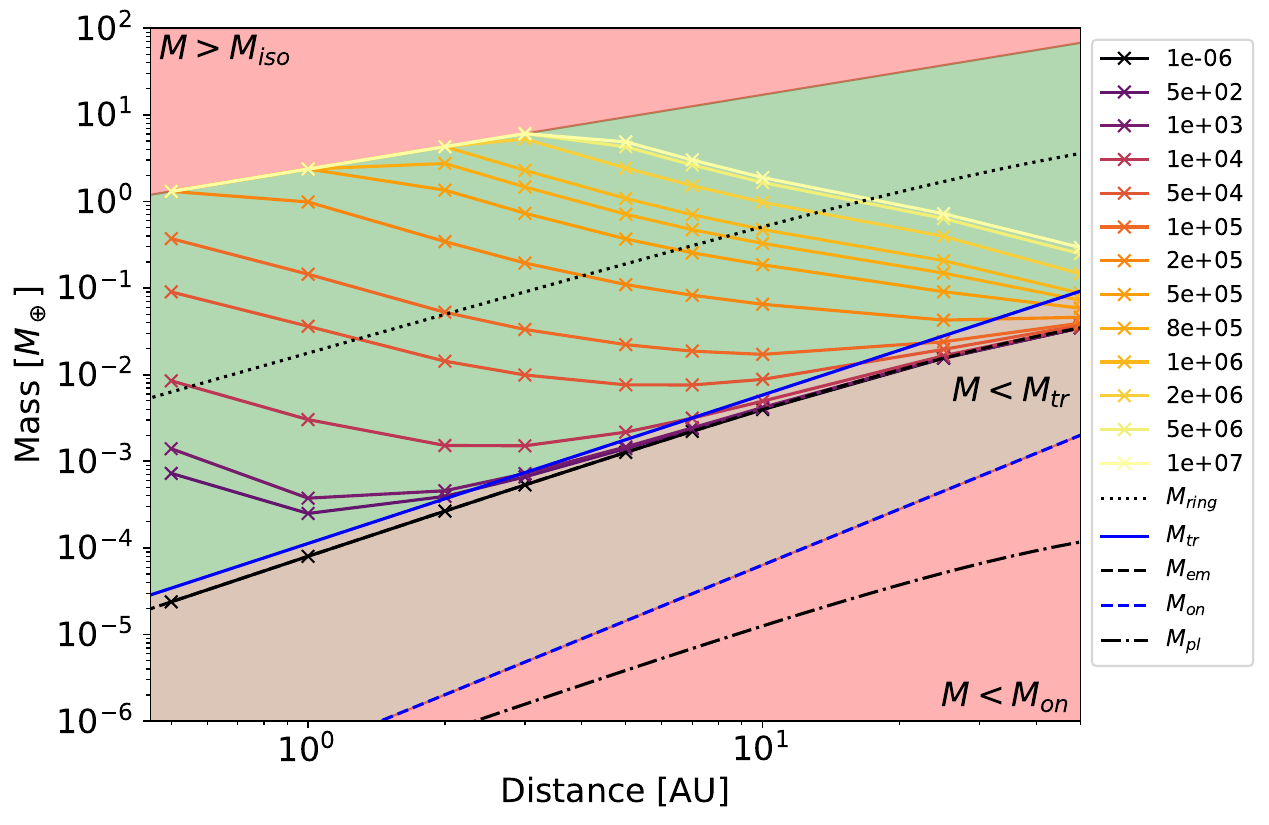}
    \caption{Same as Fig \ref{fig:M_sma_nom} but including the accretion of pebbles of $\tau_{peb} =0.1$ onto the embryo}
    \label{fig:M_sma_pebble_t}
\end{figure}

Thus far we have ignored the fact that although the accretion of pebbles is in the less efficient Bondi regime for initial masses of bodies created by streaming instability, it still contributes to the mass growth of the larger bodies. Therefore we model the accretion of pebbles onto the embryo using the approach outlined in \cite{liu_growth_2019}, considering a radially constant flux of pebbles with a fixed Stokes number of $\tau_s =0.1$. The pebble flux is calculated using the approach of \cite{lambrechts_forming_2014}. The resulting pebble flux along with the initial time $t_0$ of the different rings due to their different growth timescales and the times it takes for the planetesimals to collapse can be seen in Fig. \ref{fig:flux_peb_t}. We ignore the pebble accretion on the smaller planetesimals as their masses are significantly lower than the onset mass for all our setups and therefore pebble accretion is suppressed. The growth of the largest body when pebble accretion is included for the filaments throughout the disk is shown in Fig. \ref{fig:M_sma_pebble_t}.

\begin{figure}
    \centering
    \includegraphics[width =\hsize]{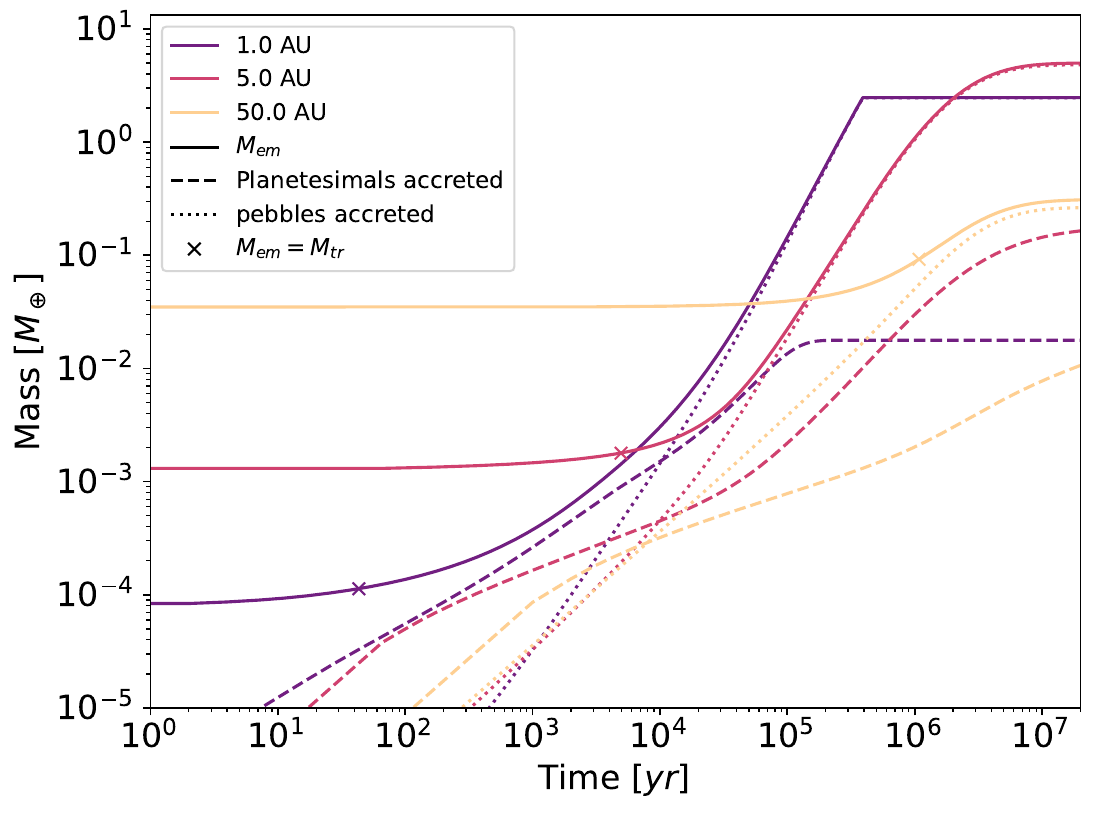}
    \caption{The total mass of the embryo (\textit{solid}), the mass budget of planetesimals accreted (\textit{dashed}) and mass of accreted pebbles (\textit{dotted})}
    \label{fig:plan_vs_pebble}
\end{figure}

We can clearly see that with the addition of pebble accretion, we are able to enhance growth significantly for filaments at all distances from the star. Up until a few au, this even allows the largest body to reach the isolation mass at which the accretion of pebbles is stopped and the rapid accretion of gas should commence. However, at larger separations beyond 25 AU, it still takes a significant amount of time for the embryo to grow and due to the decay of the pebble flux the embryo never reaches the isolation mass. To probe the growth mode in the rings in Fig. \ref{fig:plan_vs_pebble} we plot the mass of the embryos throughout time at varying locations along with the mass they accreted from planetesimals (\textit{dashed}) and pebbles (\textit{dotted}) respectively along with the time where the pebble accretion changes regime from the Bondi to the Hill regime (\textit{cross}). As expected the accretion from planetesimals only plays a role in the very beginning and gets quickly overtaken by the pebble accretion.
We see that for all separations the planetesimals are the main growth mode only for very early times and pebble accretion quickly becomes the dominant source of accretion for the largest body. This clearly shows that even for this early growth phase the accretion of pebbles can not be neglected. 

\begin{figure}
    \centering
        \includegraphics[width=\hsize]{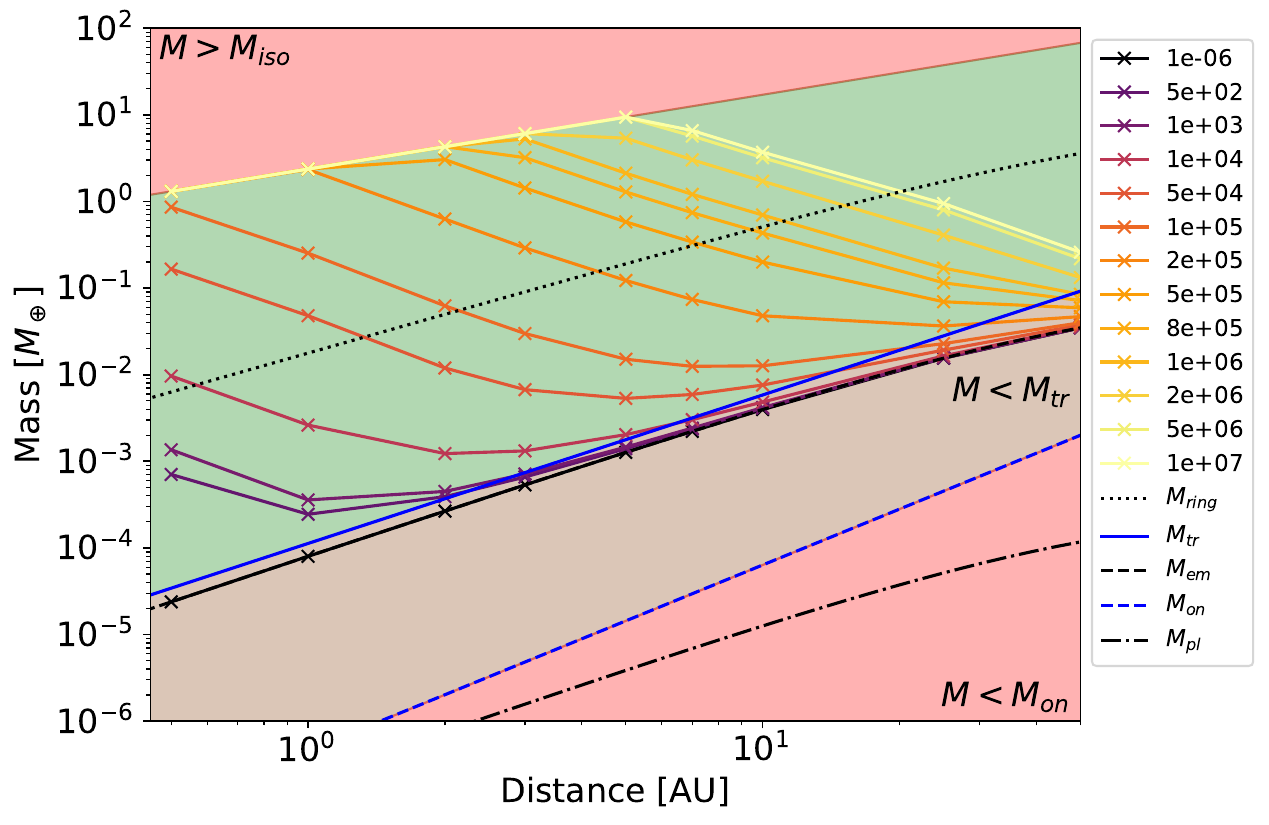}
    \centering
        \includegraphics[width=\hsize]{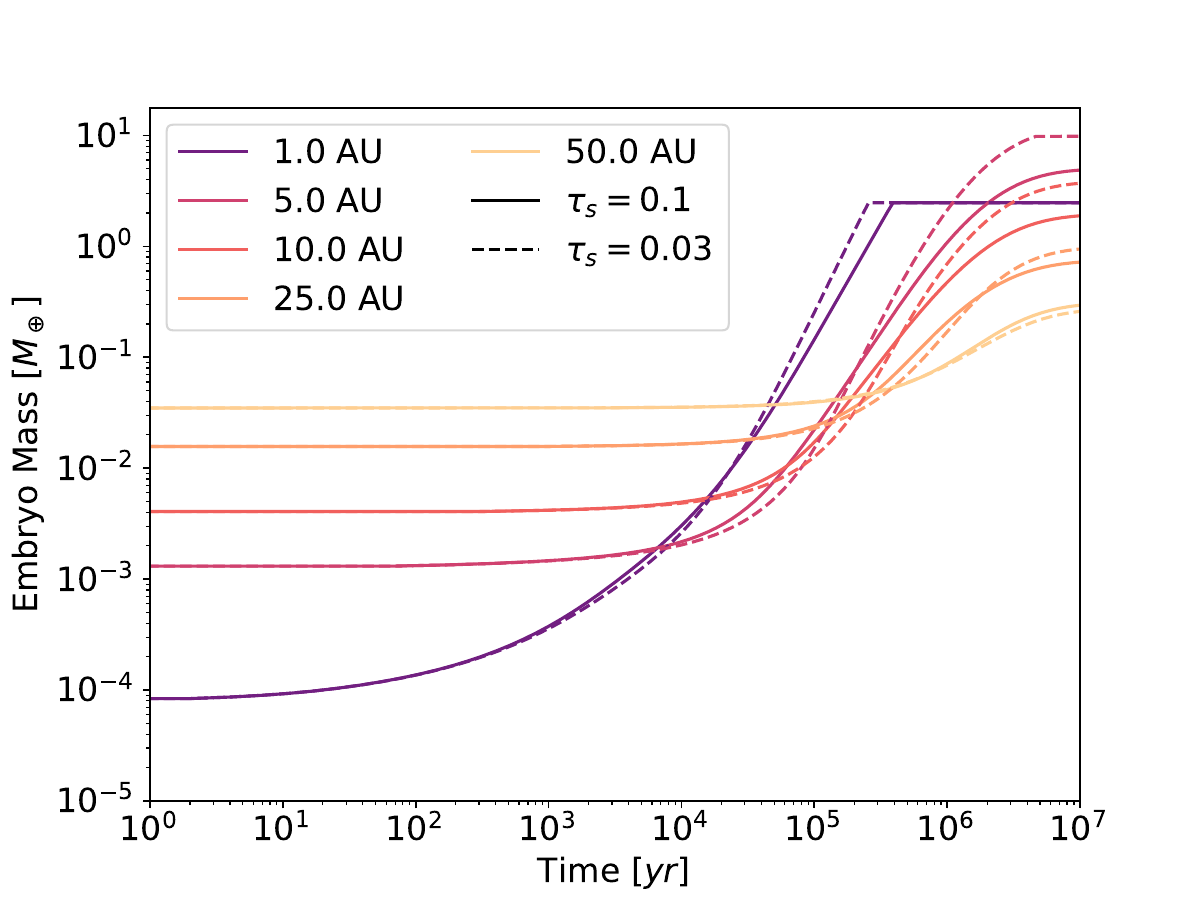}
    \caption{Same as Fig. \ref{fig:M_sma_pebble_t} for pebbles with a Stokes number of $\tau_{s} =0.03$ (\textit{top}). The mass of the embryo (\textit{bottom}) considering the accretion of pebbles with $\tau_s =0.1$ (\textit{solid}) and $\tau_s =0.03$ (\textit{dashed})}
    \label{fig:M_sma_peb_var}
\end{figure}

To probe the influence the dynamical size of the pebbles has on the early growth, we ran the same set of simulations as mentioned before but considering pebbles of a fixed lower stokes number of $\tau_{s} =0.03$. We show the results of these simulations in Fig. \ref{fig:M_sma_peb_var} along with a comparison with the growth tracks considering the accretion of pebbles with different aerodynamic sizes. We see that initially the (aerodynamically) larger pebbles result in higher accretion rates, however at higher embryo masses this trend reverses, leading to faster growth from smaller pebbles. In the inner disk, the embryo reaches the isolation mass at earlier times for the smaller pebbles and in the outer disk, this results in higher final masses. The results show that overall the early growth is enhanced for pebbles of these lower stokes numbers for the same pebble flux. To check how consistent the choice of pebbles with a fixed Stokes number is in the drift size limit, we compute the Stokes numbers in the local drift limit, following the approach of \cite{izidoro_formation_2021} which yielded Stokes numbers of $\tau_s \sim 0.01-0.07$. This motivated us to explore this second set of simulations with a fixed lower Stokes number.

\subsection{The timing of core formation}
As many global planet formation models start with partially assembled cores as their initial conditions \citep{emsenhuber_new_2021,liu_super-earth_2019,savvidou_there_2024,venturini_most_2020,venturini_fading_2024} it is important to constrain the timing of when bodies of that size are able to form. Therefore, an interesting metric to investigate at what time a filament can produce the seed for efficient pebble accretion is to check the time at which the largest body reaches the transition mass as described in Eq. \eqref{eq:M_tr} or if an embryo of that mass can even form out of these planetesimal rings. Therefore, we calculate the transition time $M_{em}(t_{tr})=M_{tr}$ i.e. the time it takes for the single largest planetesimal to grow to the transition mass for our different aforementioned setups, including the simulations considering pebble accretion and ones without. In addition to the simulations with a time-dependent pebble flux we also perform a set of simulations but considering a pebble flux that is fixed in the time of $F_{peb} = 50 M_\oplus/\text{Myr}$ for reference.

\begin{figure}
    \centering
    \includegraphics[width = \hsize]{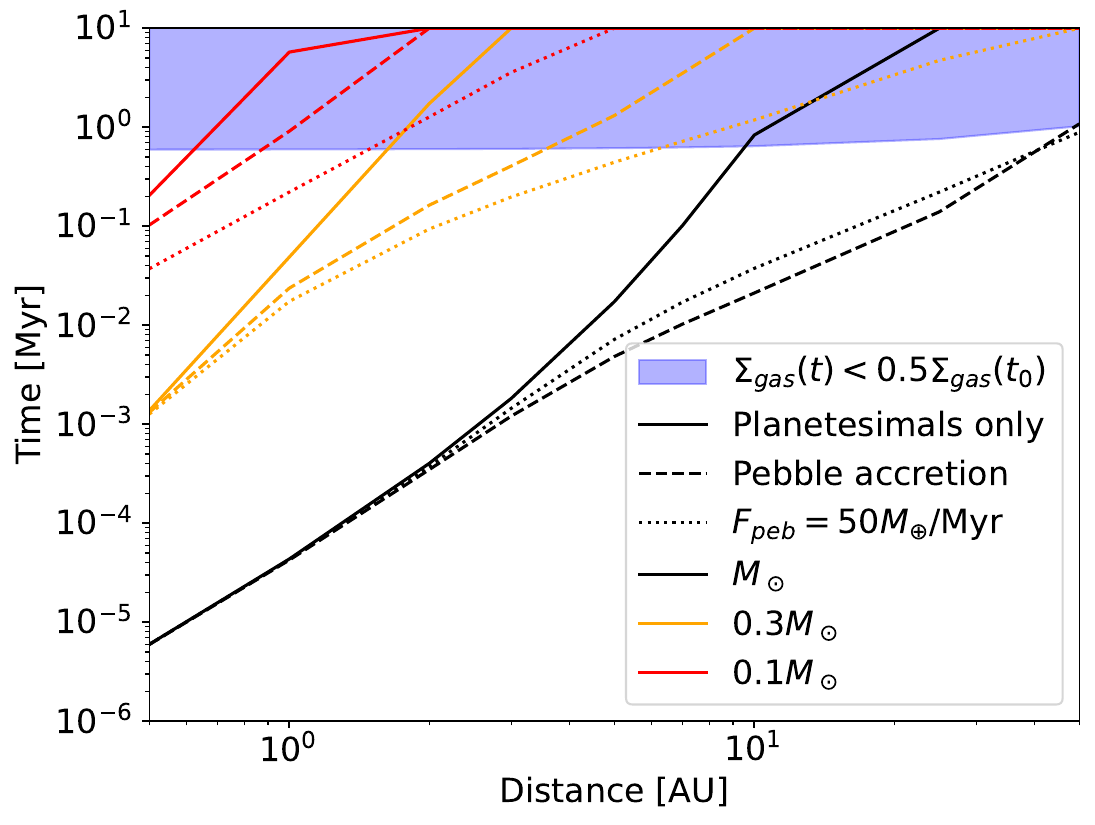}
    \caption{The time of the embryo reaching the transition mass for the nominal simulations without pebble accretion (\textit{solid}), when considering the time-dependent pebble flux (\textit{dashed}) and the fixed pebble flux (\textit{dotted}) for different stellar masses (\textit{colour}). The blue-shaded region refers to the time after which the gas surface density is reduced by half}
    \label{fig:T_trans}
\end{figure}

In Fig. \ref{fig:T_trans} we plot the time in the simulation at which the largest body reaches $M_{tr}$ as described in Eq. \eqref{eq:M_tr} to better understand the timing of embryo formation, for the simulations only considering planetesimal accretion (\textit{solid}), the time-dependent pebble flux (\textit{dashed}) and the constant pebble flux (\textit{dotted}) and for the different stellar masses (\textit{colour}). We also highlight (\textit{blue shaded region}) the time at which the gas surface density has reduced by half of the initial value as a proxy for the disk lifetime as we do not consider external photoevaporation or other mass loss mechanisms for our gas disk. As we can clearly see the time needed to reach the transition mass monotonically increases with distance to the star and with decreasing stellar mass. We can also clearly see that the addition of pebble accretion reduces the time to reach the transition significantly although the effect is more visible for the rings at larger separations as in the inner disk there is still a significant contribution from planetesimal accretion. 

\begin{figure}
    \centering
    \includegraphics[width=\hsize]{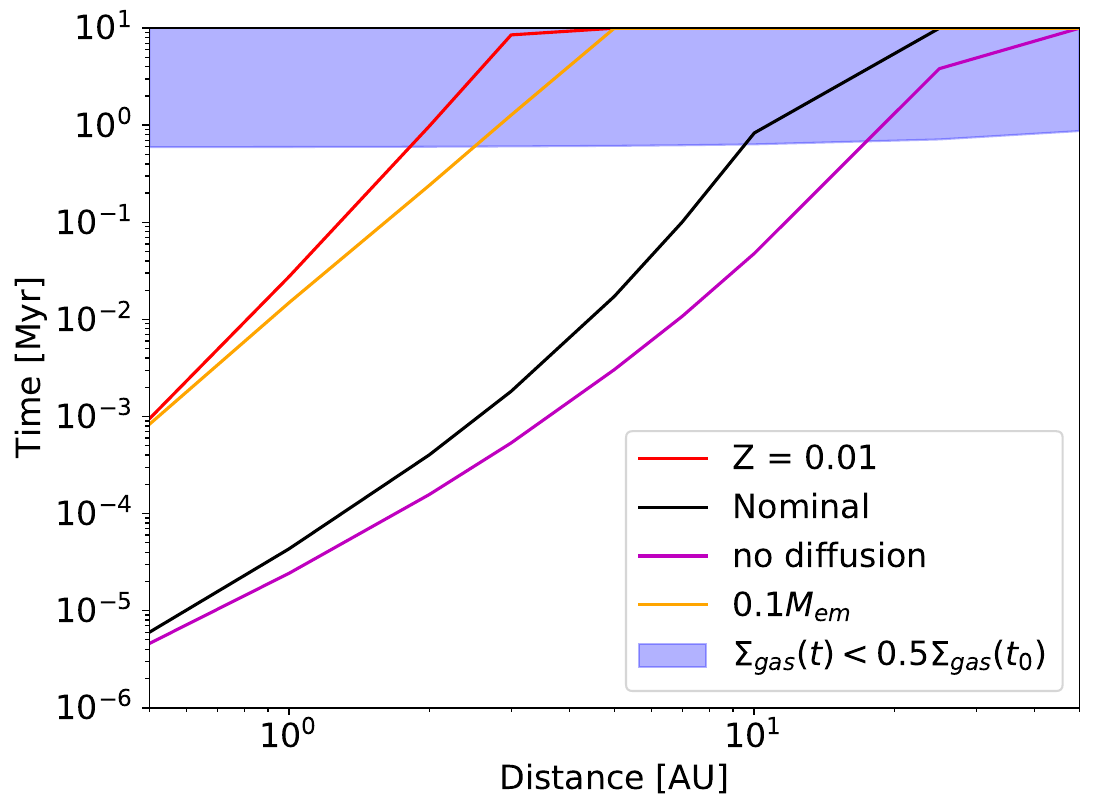}
    \caption{Same as Fig. \ref{fig:T_trans} for the following setups, nominal (\textit{black}), with a reduced initial embryo mass (\textit{yellow}), reduced filament metallicity (\textit{red}) and ignoring diffusion (\textit{magenta})}
    \label{fig:T_trans_var}
\end{figure}

Additionally, to probe the influence of the initial mass function of the planetesimals and the properties of the initial filament we perform another set of simulations considering a reduced filament metallicity of $Z = 0.01$ (\textit{red}) and one where we reduced the initial mass of the largest planetesimal by a factor of 10 while keeping the total mass constant (\textit{yellow}). The results of this investigation can be seen in Fig. \ref{fig:T_trans_var} where we once again plot the time for the largest planetesimal to reach the transition mass.

 As we expect, the reduction in initial embryo mass leads to a significantly longer formation timescale so that outside of a few AU it is not able to reach the transition mass at all. The simulations with the reduced metallicity show a very similar picture with even slightly longer timescales as in addition to a reduced initial embryo mass, the surface density of planetesimals is also reduced leading to longer accretion timescales. 

\begin{figure}
    \centering
    
    \includegraphics[width=\hsize]{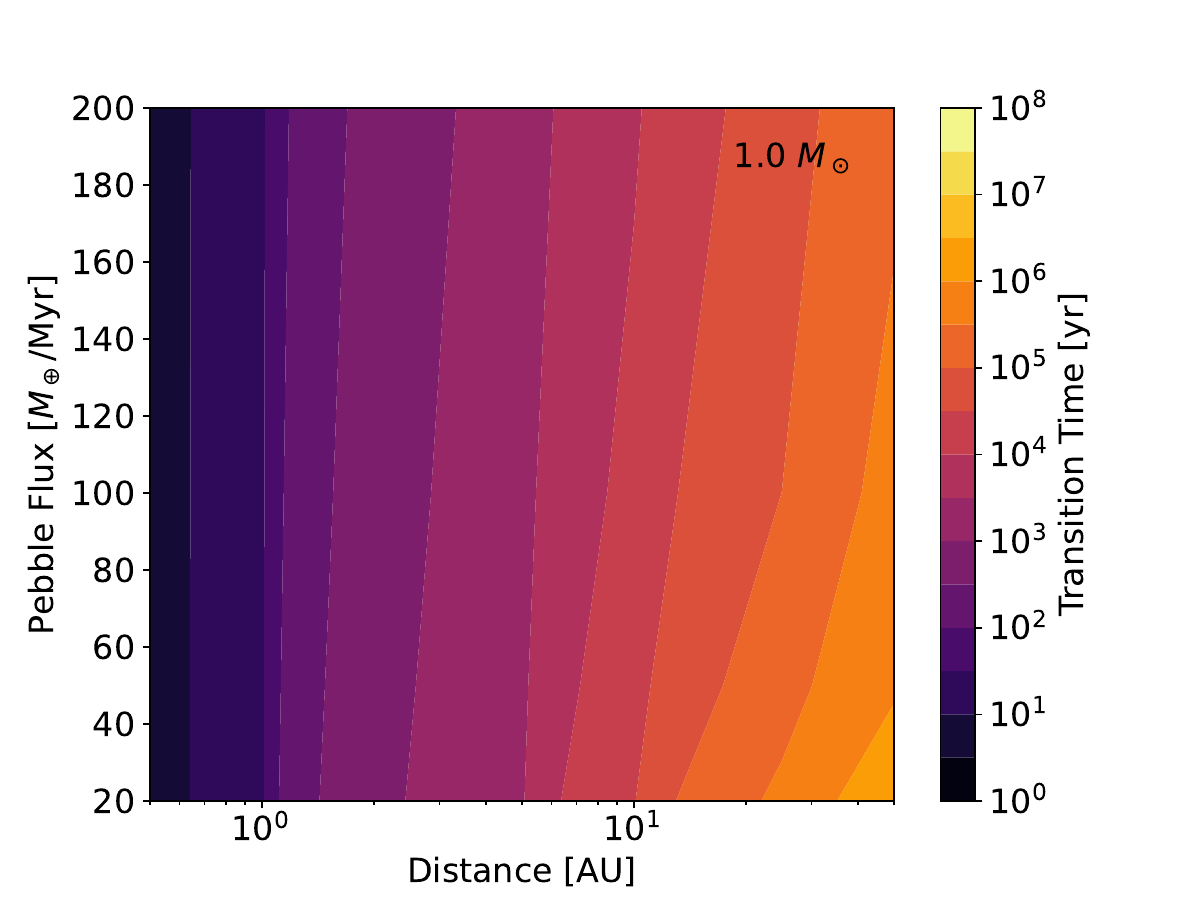}
    \includegraphics[width=\hsize]{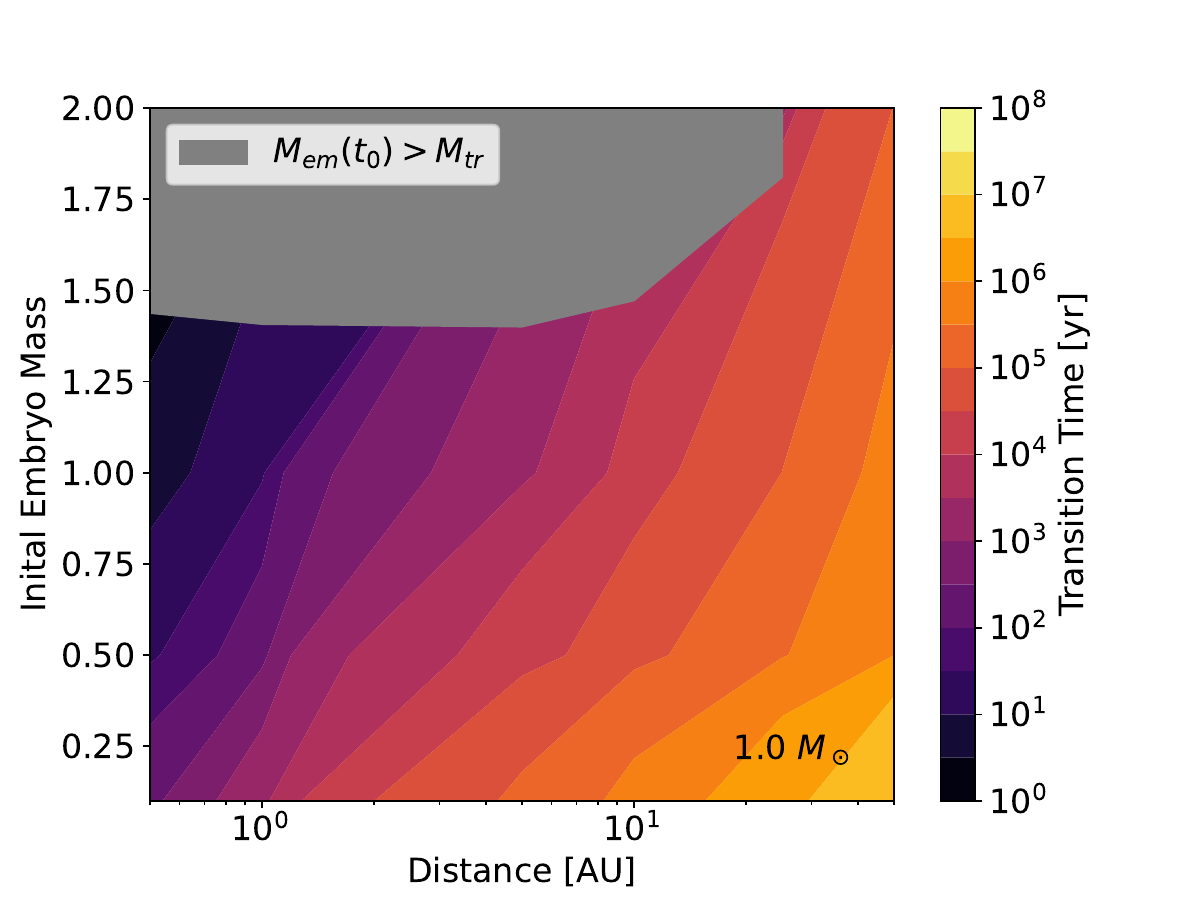}
    \includegraphics[width=\hsize]{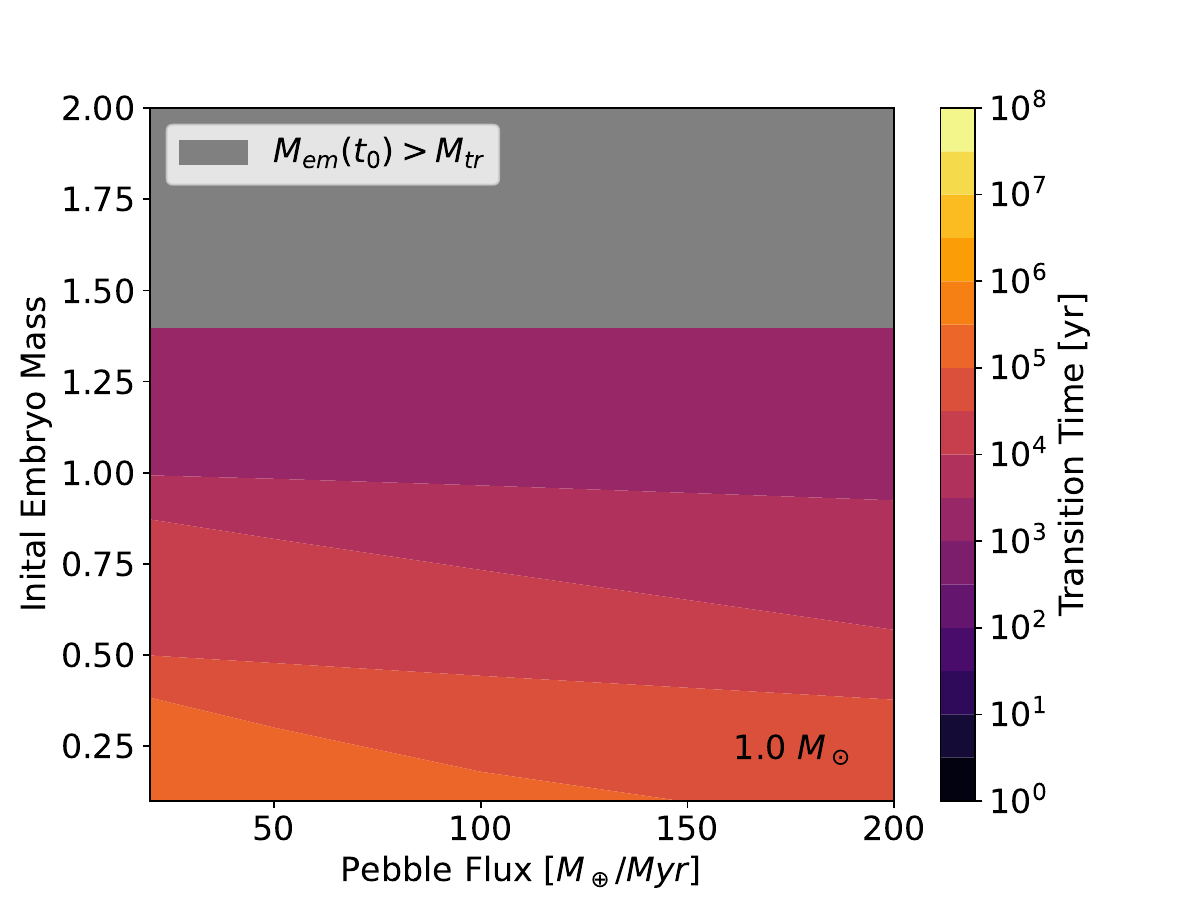}
    
    \caption{The transition timing (\textit{colour}) for the grid simulations around a $1M_\odot$ mass star as a function of two gird dimensions where the last variable not plotted is given by the bold value in Table \ref{tab:grid_params}. The grey area refers to simulations where the initial embryo is already larger than the transition mass}
    \label{fig:grid_plot}
\end{figure}

\begin{figure}
    \centering
    \includegraphics[width=\hsize]{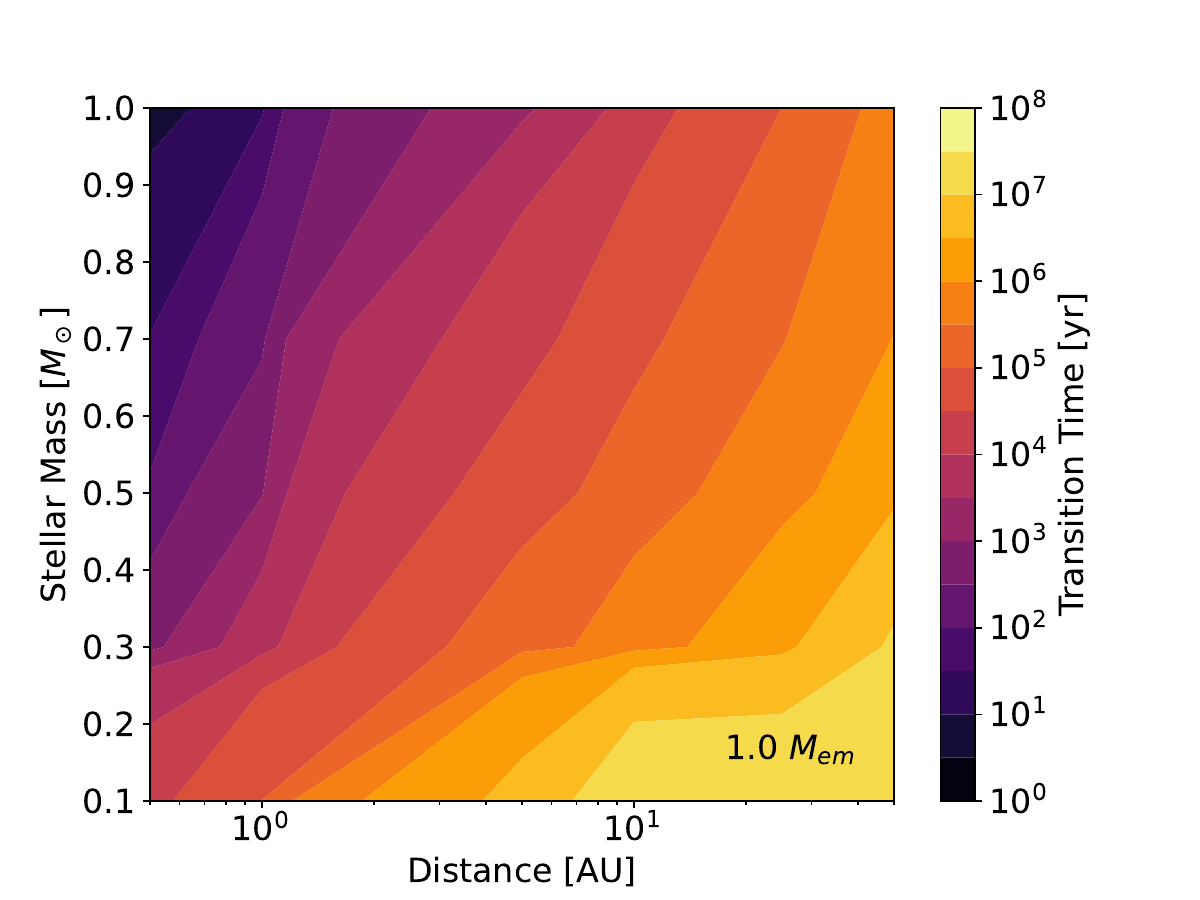}
    
    \caption{The transition timing of for the grid at different stellar masses and separations where $F_{peb} = 100 M_\oplus/\rm{Myr}$ and $M_{em} =1M_{em,0}$}
    \label{fig:grid_stellar}
\end{figure}

In order to give an intuition on what model parameters influence the timing of embryo formation the most and to give a simple description that can inform population synthesis models at what time to insert the initial partially assembled embryos in their simulations we perform a grid of simulations, varying multiple key model parameters: the semi-major axis of the filament, the stellar mass, the pebble flux and the mass of the largest embryo. The timing of embryo formation is inferred by calculating the transition timing for each simulation. In Table \ref{tab:grid_params} we present the grid chosen for the different parameters along with the nominal value (\textit{bold}) can be found in Table \ref{tab:grid_params}. Note we varied the initial embryo mass as a factor of the nominal model i.e. $M_{em,grid} = f_{em}*M_{em,0}$ rather than choosing a fixed value to keep the investigated scenario consistent with the planetesimal IMF we consider. A visualisation of simulation results can be seen in Fig. \ref{fig:grid_plot}, where for a fixed stellar mass, we plot the transition timing for the remaining parameter pairs. In the interest of brevity, the same plots for further stellar masses can be found in Figure \ref{fig:grid_plots} in the Appendix. Additionally, we plot the transition timing for different stellar masses at different distances to show the influence of the stellar mass on core formation in Figure \ref{fig:grid_stellar}.

As expected, we recover the observed behaviour of the effects of the distance from the star and the pebble flux on the transition timing. We also see that the effect of the distance is much more important for the formation time when compared to the pebble flux whose influence is lower, especially for the solar mass star and in the inner disk. The initial embryo mass is highly important for the subsequent growth and its effect dominates over the pebble flux but is less influential around lower-mass stars (as can be seen in Figure \ref{fig:grid_plots}) where the pebble flux gains more importance. This is consistent with the simulations with higher metallicity we ran for low-mass stars that did not enhance growth significantly.
To make the data easily accessible it has been uploaded as a Python package to GitHub \footnote{https://github.com/NIcolas-Kaufmann/emerge}, which contains the transit timings for all the simulations along with the time (if) the embryo reaches $10^{-3}$ and $10^{-2} M_\oplus$. We note that even though the calculated initial times are model-dependent and therefore should be treated as an estimate, due to the modular nature of the model, it can be easily adapted to various scenarios by for example changing disk and planetesimal initial mass function which we aim to explore in future works.

\begin{table}[]
    \centering
    \begin{tabular}{c|c}
    Parameter & Value  \\
    \hline
        Distance [AU]&  [0.5, 1, \textbf{5}, 10, 25, 50 ]\\
        Stellar mass $[M_\odot]$ & [\textbf{1}, 0.7, 0.5, 0.3, 0.1]\\
        Pebble flux $[M_\oplus/\text{Myr}]$& [200, \textbf{100}, 50, 20]\\
        Mass factor $f_{em}$ & [2, \textbf{1},0.5, 0.1]\\
        
    \end{tabular}
    \caption{Parameter gird for the simulations to infer the timing of embryo formation}
    \label{tab:grid_params}
\end{table}

\section{Discussion and conclusions}
\label{sec:conlusions}
\subsection{Limitations of the model}
There are several simplifications that we made for the model described in this paper that have to be discussed. Firstly we neglected the migration of both the embryo and the planetesimals. However, in the mass regime of the embryo, we describe in this paper that the migration timescales are much longer than the time it takes to reach the transition masses, nevertheless for the cases where the embryo grows larger, up to its isolation mass, these effects should be considered \citep{paardekooper_torque_2011,ida_new_2020}.
Due to the fact that the torque generated by the disk onto the planet scales with the planet's mass until the embryos reach the transition mass, planet migration should not play a relevant role \citep{tanaka_three-dimensional_2002, paardekooper_torque_2011, ida_new_2020}. However, when the mass of the embryo exceeds the transition mass planet migration could be too crucial to be ignored. In addition, if pebble accretion becomes very efficient, thermal torques \citep{guilera_importance_2021, baumann_influence_2020} or dust torques \citep{guilera_quantifying_2023} could be relevant as well. 

Regarding the planetesimals, for those of initial sizes $r_p > 10$ km, the drift is negligible \citep{ormel_understanding_2012}, and only for planetesimals of radii $\lesssim 1$~km planetesimal drift plays an important role \citep{guilera_consequences_2010, guilera_giant_2017}. For the fragments produced in collisions, the radial transport is significant leading to further depletion of solids in the ring so our simulations can be seen as upper limit when we include the fragmentation of planetesimals. The choice of the initial mass function of the planetesimals significantly shapes the nature of the subsequent growth in the ring \citep{liu_growth_2019}. However, as planetesimal formation is a very active field of research, the precise nature of the initial mass function remains very uncertain both from a formation point \citep{polak_high_2022, schafer_initial_2017} and also from observational constraints \citep{morbidelli_asteroids_2009,schlichting_initial_2013} from the solar system. Therefore we limit ourselves to the IMF as described in Section \ref{sec:initial}. Additionally, our model only considers the formation in a single ring whereas a fully global model could model the formation of many concurrently forming planets and rings leading to interactions between them for example by reducing the pebble flux for further filaments downstream. 

For this study, we chose simple dust evolution and pebble accretion models, as this makes it easier to investigate the influence of the different model parameters (e.g. the pebble flux or stokes number) on the early growth stage. However, as a trade-off, there are certain characteristics of pebble accretion we have to neglect. Firstly, even though we infer the pebble flux from the disk properties \citep{lambrechts_forming_2014}, we do not calculate the dust and pebble properties self-consistently throughout the disk as could be done by modelling the dust evolution using models of varying complexity (e.g. \citealp{drazkowska_how_2021,pfeil_tripod_2024,stammler_dustpy_2022}). This would then also allow us to account for the shift in accretion efficiencies due to the presence of pebbles of different sizes. It was shown by \cite{lyra_analytical_2023} that accounting for the size distribution of pebbles significantly changes the accretion efficiency in the Bondi regime. This would promote enhanced early growth i.e. it would reduce the effective onset mass of pebble accretion. Using their description of poly-disperse pebble accretion they show that the initial embryos in the same mass range as considered in this work show significantly lower accretion timescales compared to single-size pebbles. The higher accretion rates due to the presence of pebbles of smaller sizes were qualitatively shown in Fig. \ref{fig:M_sma_peb_var} but to properly account for the poly-disperse nature of the pebble flux is beyond the scope of this work (and incompatible with the analytic calculation of the pebble flux) and therefore will be explored in future works.
\subsection{Conclusions and summary}

In this paper, we investigated the early growth of a planetary embryo from a ring of planetesimals to sizes large enough to accrete pebbles efficiently. We developed a formation model that tracks the evolution of a ring of planetesimals created from streaming instability. We simulate the planetesimals and the largest body including the relevant physics to evolve the system self consistently considering the evolution of the random velocities mutual collisions and pebble accretion. Our main findings are:

\begin{itemize}
    \item Growing the seed of pebble accretion from a planetesimal ring, when only considering the growth from planetesimal accretion, is possible only in the inner disk ($< 1AU$) on short timescales. However, the embryo growth from planetesimal accretion is negligible at large separations.
    \item The diffusion of the planetesimal ring is a major inhibitor for the growth of the largest body.
    \item The inclusion of pebble accretion can not be neglected as it contributes significantly to the growth of the embryo even in the slower Bondi regime. Still, at large separation $\approx50 AU$, it takes a considerable amount of time for the largest planetesimal to reach the transition mass from the Bondi to the Hill regime.
    \item For lower stellar masses the growth of the largest body is slower and even at moderate distances $\approx 1-10$ AU it remains hard to grow the core of planets within the typical lifetime of protoplanetary disks. These results could be related to the lack of giant planets compared to more massive stars \citep{sabotta_carmenes_2021}.
    \item The timing of core formation is strongly dependent on stellar mass and semi-major axis which should be taken into account for formation models starting with partially assembled cores.
\end{itemize}

Understanding the formation pathway presented in this work and in similar works \citep{liu_growth_2019,jiang_efficient_2022,lorek_growing_2022} is vital to constrain the early stages of planet formation and the interplay planetesimal formation and planetesimal/pebble accretion. However, these works all consider the isolated formation of planets at a single location. This calls for the investigation of these formation scenarios using global evolution models allowing for the formation of planetesimals in multiple rings and the interaction among them \citep{lau_sequential_2024}. Furthermore, it highlights the necessity to better constrain the IMF of planetesimals as it has a major impact on the subsequent growth in these filaments.

\begin{acknowledgements}
O.M.G. gratefully acknowledges the invitation and financial support from the International Space Science Institute (ISSI) Bern in early 2024. This research was partially supported through the Visiting Scientist program of the ISSI in Bern. O.M.G. and I.L.S.S. are partially supported by PIP-2971 from CONICET (Argentina) and by PICT 2020-03316 from Agencia I+D+i (Argentina). We acknowledge the support from the NCCR PlanetS
\end{acknowledgements}

\bibliographystyle{bibtex/aa}
\bibliography{bibtex/ref_2.bib}
\begin{appendix}
\section{Validation and comparison of the fragmentation model}
\label{sec:app_frag}

\begin{figure}
    \centering
    \includegraphics[width=\hsize]{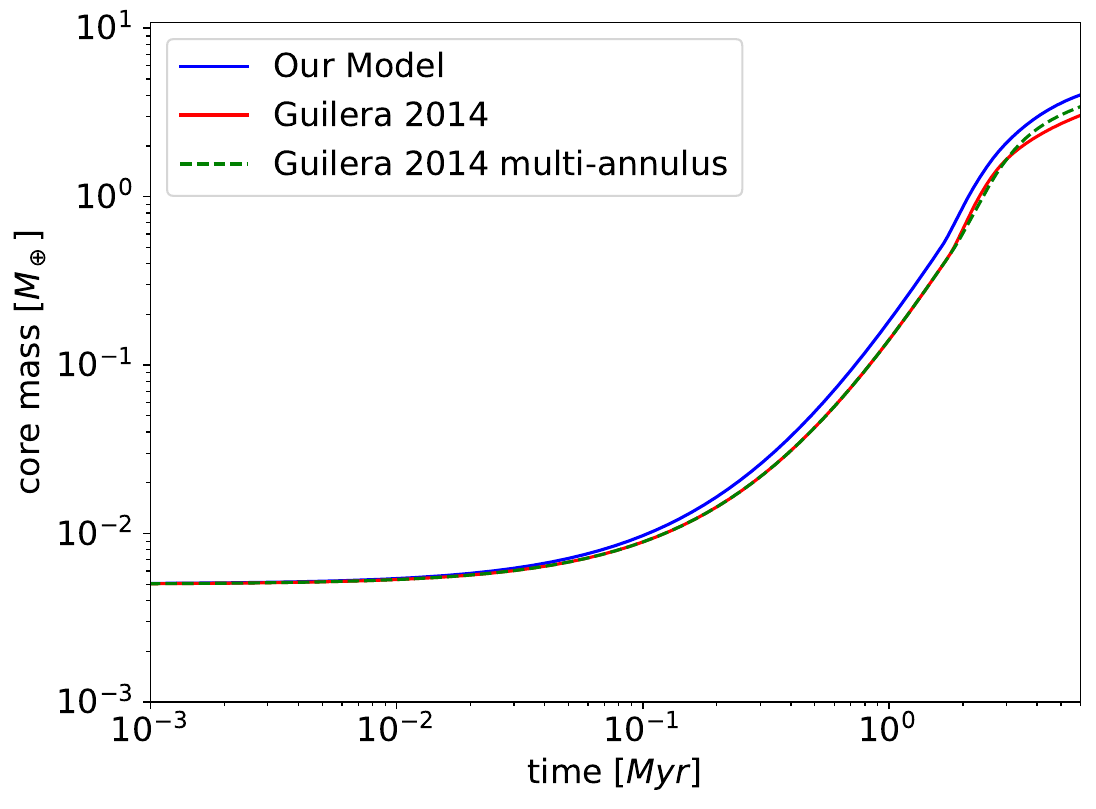}
    \caption{The growth track of a moon-mass embryo at 5 AU in our model (\textit{blue}) and \citep{guilera_planetesimal_2014} with only the local collisions (\textit{red}) and the multi-annulus case (\textit{green})}
    \label{fig:quilera_track}
\end{figure}

In order to test the validity of the simplifications made in the fragmentation model derived in this work we compare it against the model implemented in \citet{guilera_planetesimal_2014} and \citet{sebastian_planetesimal_2019}. To do this, and to isolate the effects of the fragmentation description the treatment of the remaining physics is kept as simple as possible to perform the comparison. We compute the formation of a planet just considering the accretion of planetesimals of $100$~km radius and the fragments generated by the planetesimal fragmentation process, i.e. we do not consider gas accretion onto the planet or the enhancement in the planet's cross-section of the capture radius due to the presence of an envelope. We consider the same initial conditions as in \cite{guilera_planetesimal_2014} and \cite{sebastian_planetesimal_2019}, i.e. the growth of an initial mass Moon embryo located at 5 AU and immersed in a planetesimal disk ten times more massive than the minimum mass solar nebula \citep{hayashi_structure_1981}. The gas disk is considered to decay with a characteristic timescale of $6$ Myr. Additionally, we do not consider accreting collisions among planetesimals i.e. collisions where Equation \ref{eq:M_R} $M_R > M_T$ in Eq. \eqref{eq:M_R}. We note that we only consider the fragmentation of planetesimals that belong to the same annulus \citep[as in][]{kaufmann_influence_2023}, but for the sake of comparison we also compute the multi-annulus case, i.e. where targets and projectiles can belong to different annulus. In figure \ref{fig:quilera_track} we can see the resulting growth track of the planetary embryo using the different codes.

\begin{figure}
    \centering
    \includegraphics[width = \hsize]{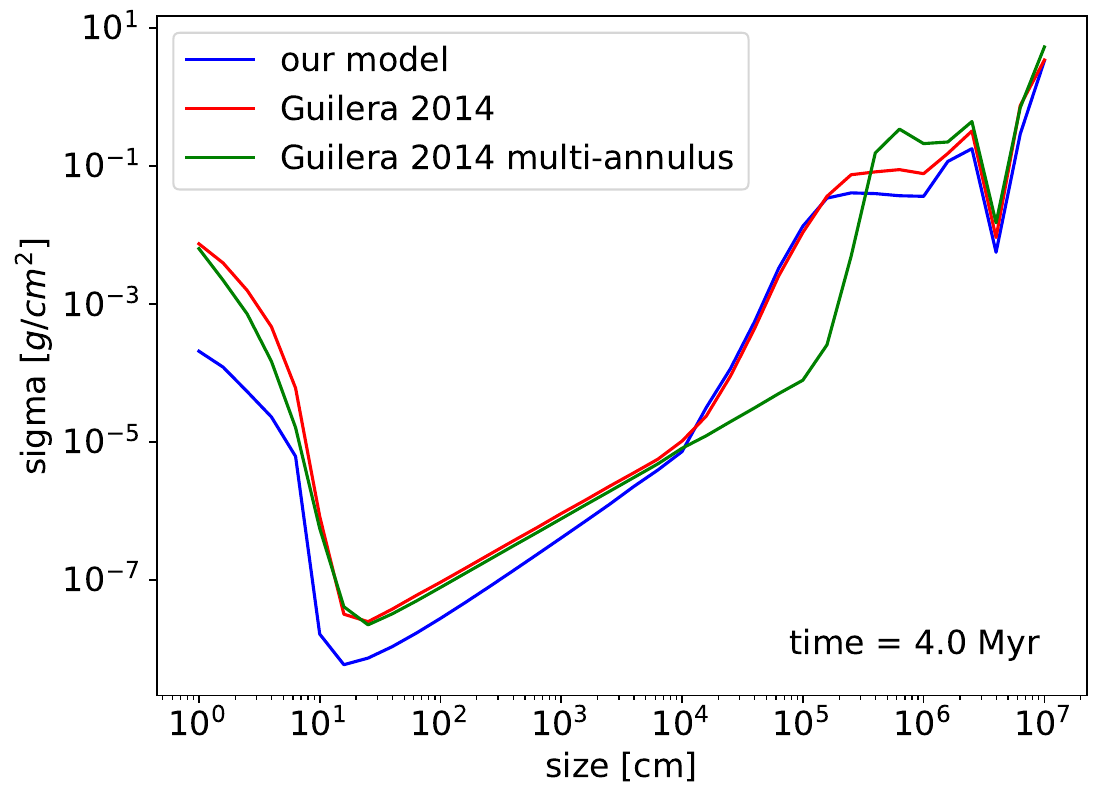}
    \caption{The surface density distribution of the planetesimals at the location of the planet calculated using our model (\textit{blue}) and the model presented in \cite{guilera_planetesimal_2014} with only the local collisions (\textit{red}) and the multi-annulus case (\textit{green})}
    \label{fig:dist_p}
\end{figure}

The growth tracks in both models are a good match. Our code seems to have slightly faster growth leading to an increased final mass which can be explained because of the way we calculate the feeding zone which differs in both codes (\cite{guilera_planetesimal_2014} considers a feeding zone with a half-width of  $4R_h$ with a soothing function whereas we consider at top with a half-width of $5 R_h$) as this difference shows up before the first fragments are created. As the focus of this comparison is the fragmentation model we are also interested in the size distribution of the planetesimals at the planet's location throughout the formation process. To illustrate those we show in Fig. \ref{fig:dist_p} the surface density distribution of the different sizes at $4$ Myr in both models. 

As we can see even though there are some differences in the size distribution they seem to fit very well overall however for the smaller sizes there seem to be some deviations stemming from the fact that we only consider local collisions and that for a given time the embryo in both simulations does not have the same mass resulting in a slight difference in the evolution of the size distribution of the planetesimals.
\section{Model parameters}

\begin{table}[h]
    \begin{tabular}{ccc}
        \hline
        Parameter & Value & Description\\
        \hline
        \hline
       $M_\star$  & $M_\odot$ & Stellar mass \\
       $Z$  &  0.1 & filament metallicity\\
       $Z_0$ & 0.01 & disk metallicity \\
       $\epsilon_d$ & 0.5 & dust growth efficiency \\
       $\tau_p$ & 0.1 & Stokes number of pebbles \\
        $p_{eff}$ &  1 & planetesimal formation  \\ & & efficiency \\
        $\rho_{s}$ &  2 g/cm & bulk density of solids \\
        $\alpha$ & $10^{-2}$ & gas turbulence parameter\\
        $r_1$ & 71,94,121 AU  & disk truncation radius\\ & & ($M_\star \in [1,0.3,0.1]M_\odot$)\\
        $v_{frag}$ & $1$ m/s & fragmentation speed \\
        $\alpha_z$ & $10^{-4}$ & mid plane turbulence\\
        \hline
        
    \end{tabular}
    \caption{Table with the chosen parameters for the simulations}
    \label{tab:nominal_params}
\end{table}
\newpage
\onecolumn
\section{Grid simulations}
\label{sec:app_grid}

\begin{figure*}[h]
    \centering
    \begin{minipage}[b]{0.45\textwidth}
        \centering
        \includegraphics[width=\textwidth]{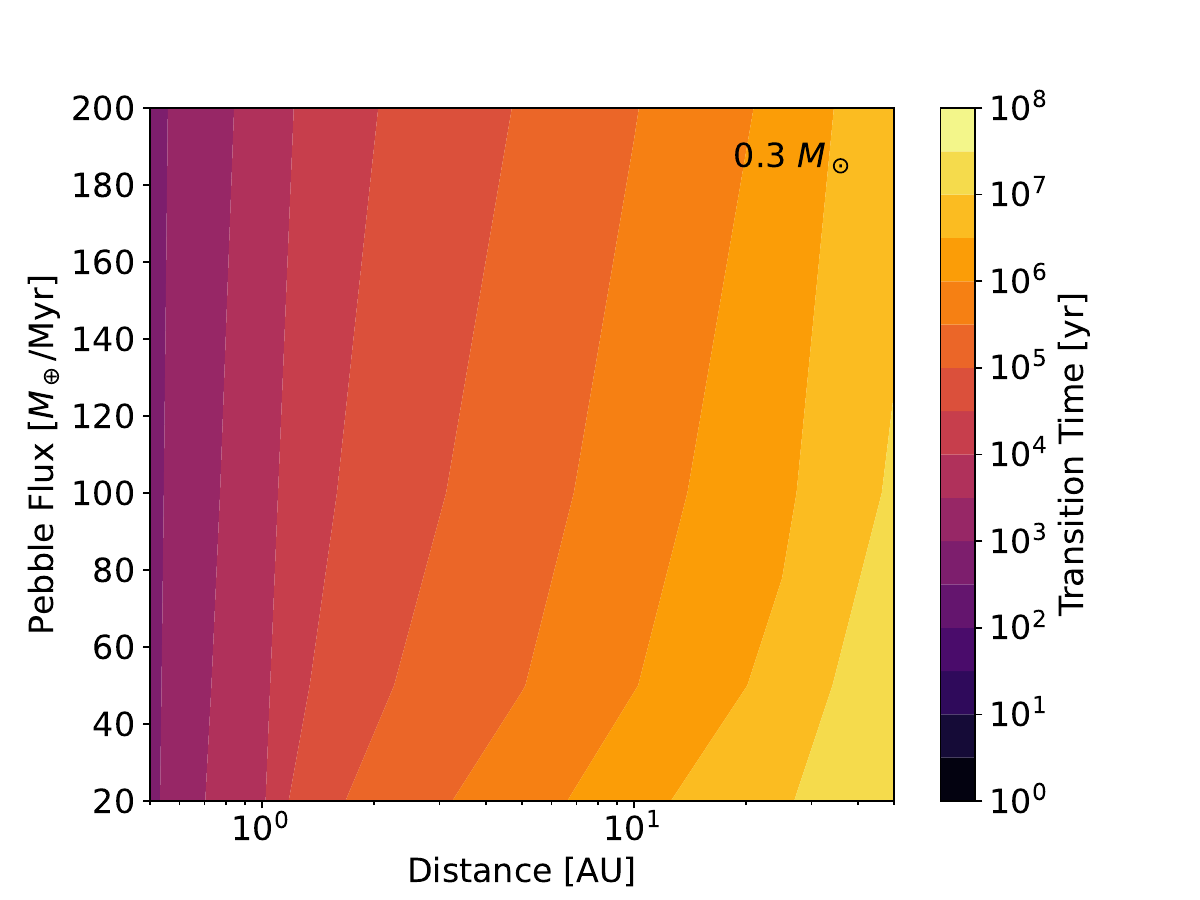}
    \end{minipage}
    \hfill
    \begin{minipage}[b]{0.45\textwidth}
        \centering
        \includegraphics[width=\textwidth]{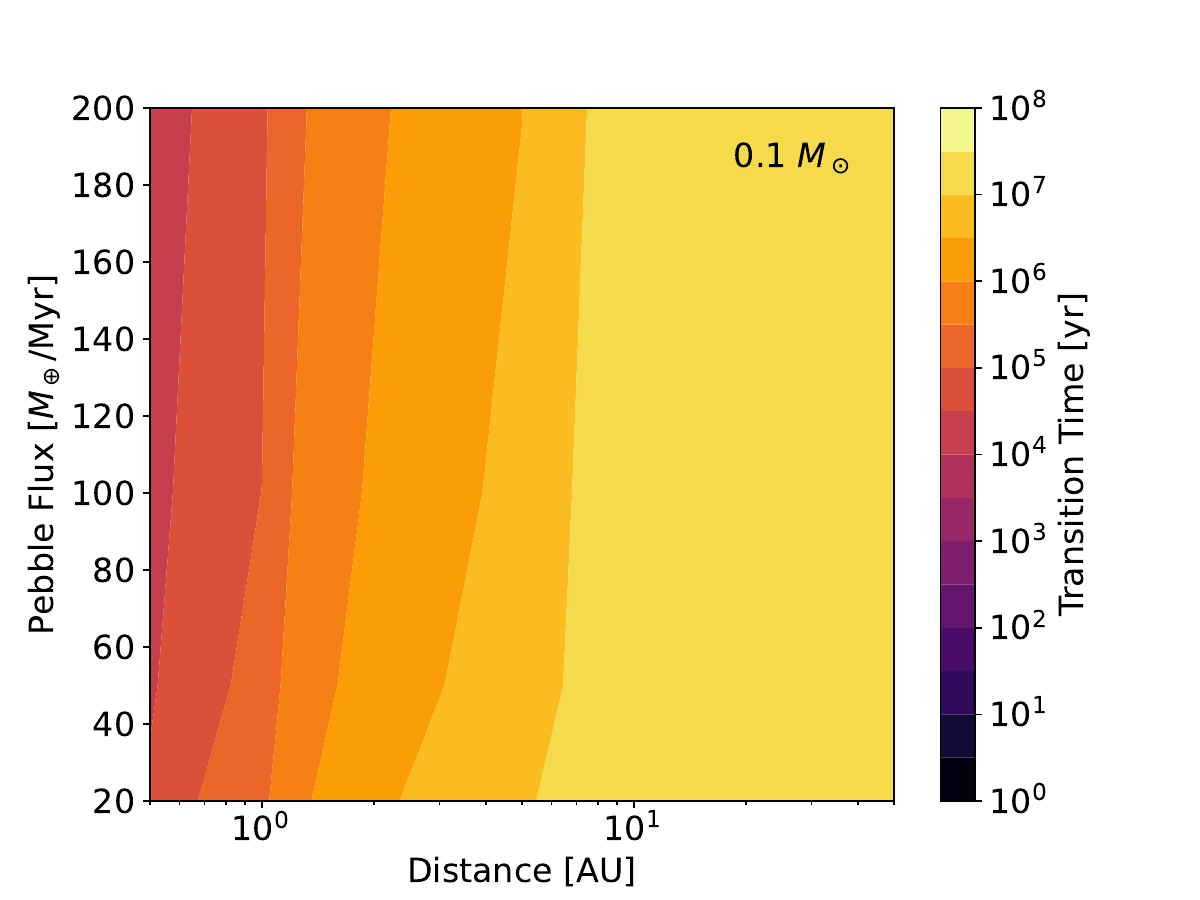}
    \end{minipage}

    \begin{minipage}[b]{0.45\textwidth}
        \centering
        \includegraphics[width=\textwidth]{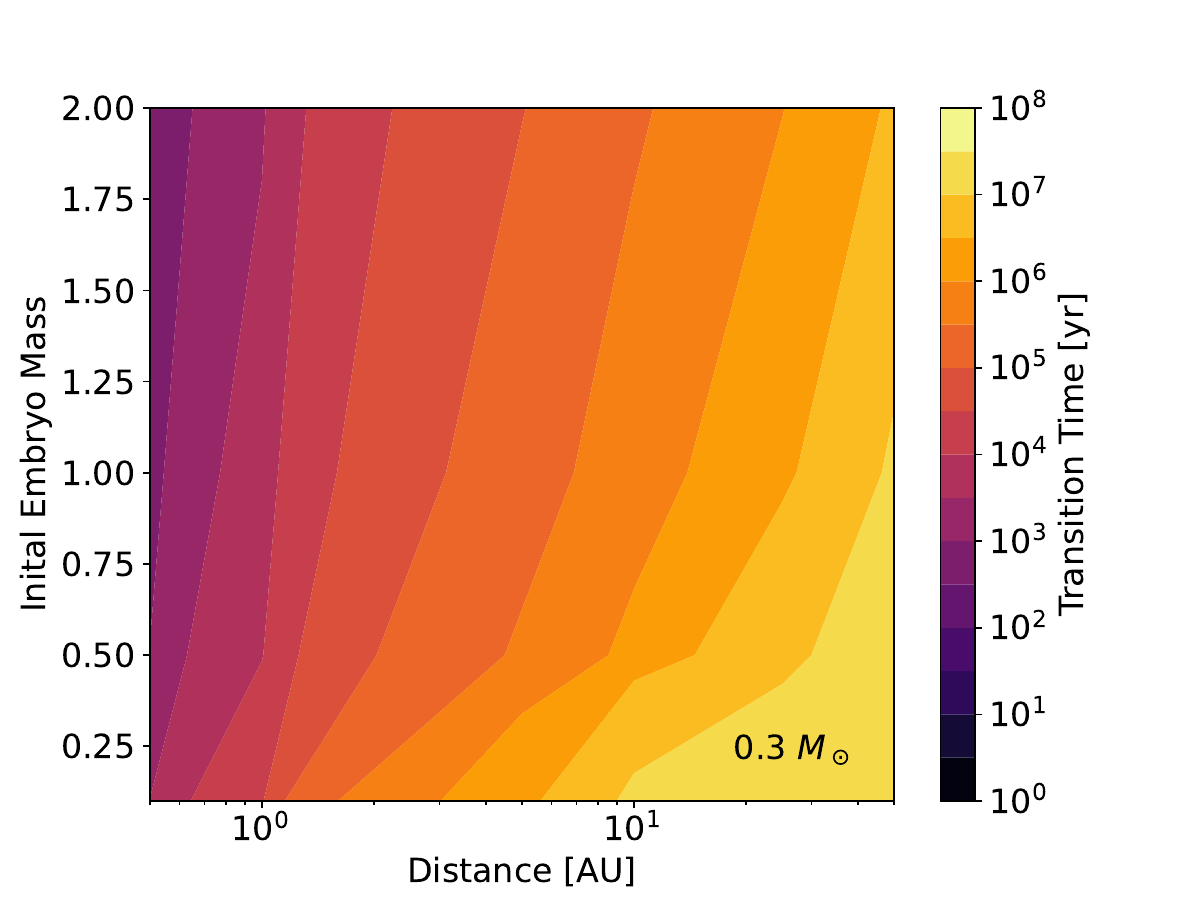}
    \end{minipage}
    \hfill
    \begin{minipage}[b]{0.45\textwidth}
        \centering
        \includegraphics[width=\textwidth]{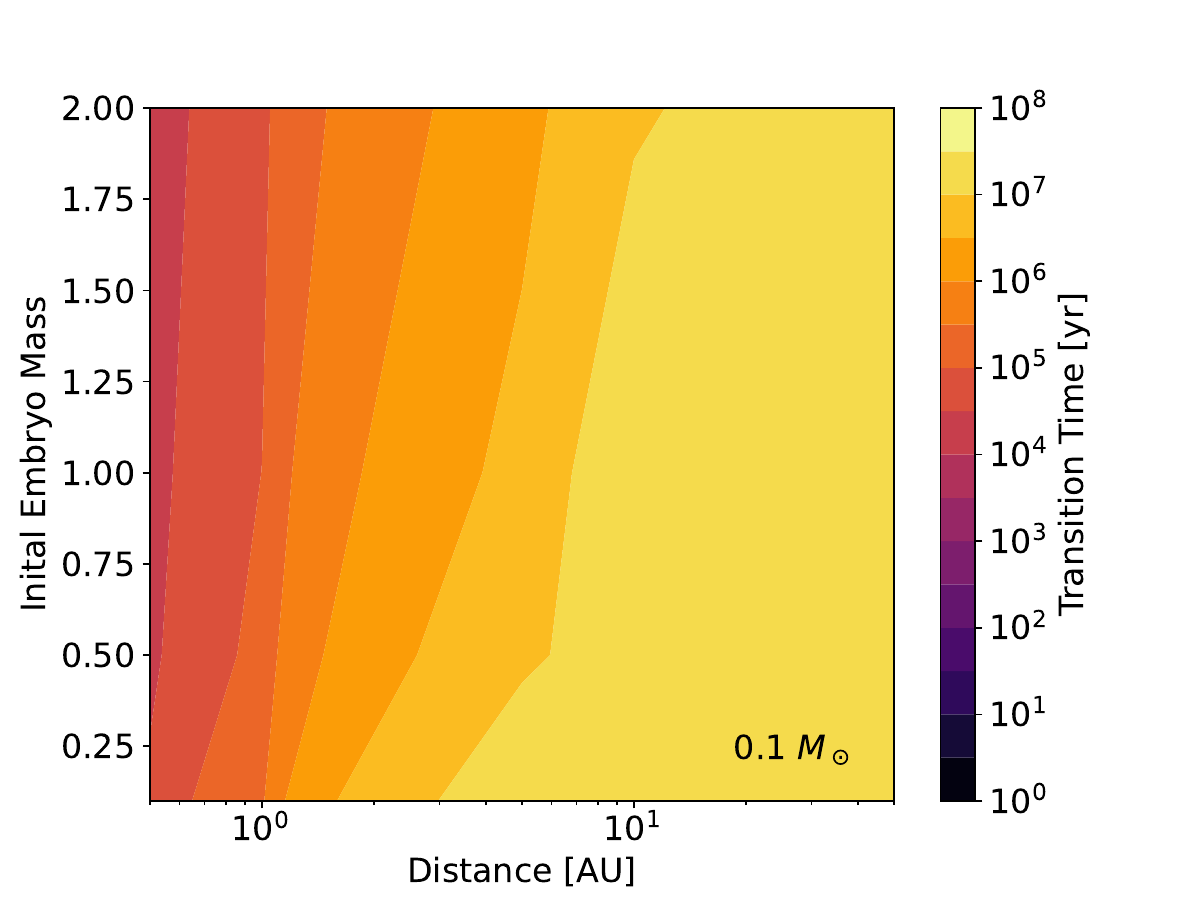}
    \end{minipage}
    
    
    \begin{minipage}[b]{0.45\textwidth}
        \centering
        \includegraphics[width=\textwidth]{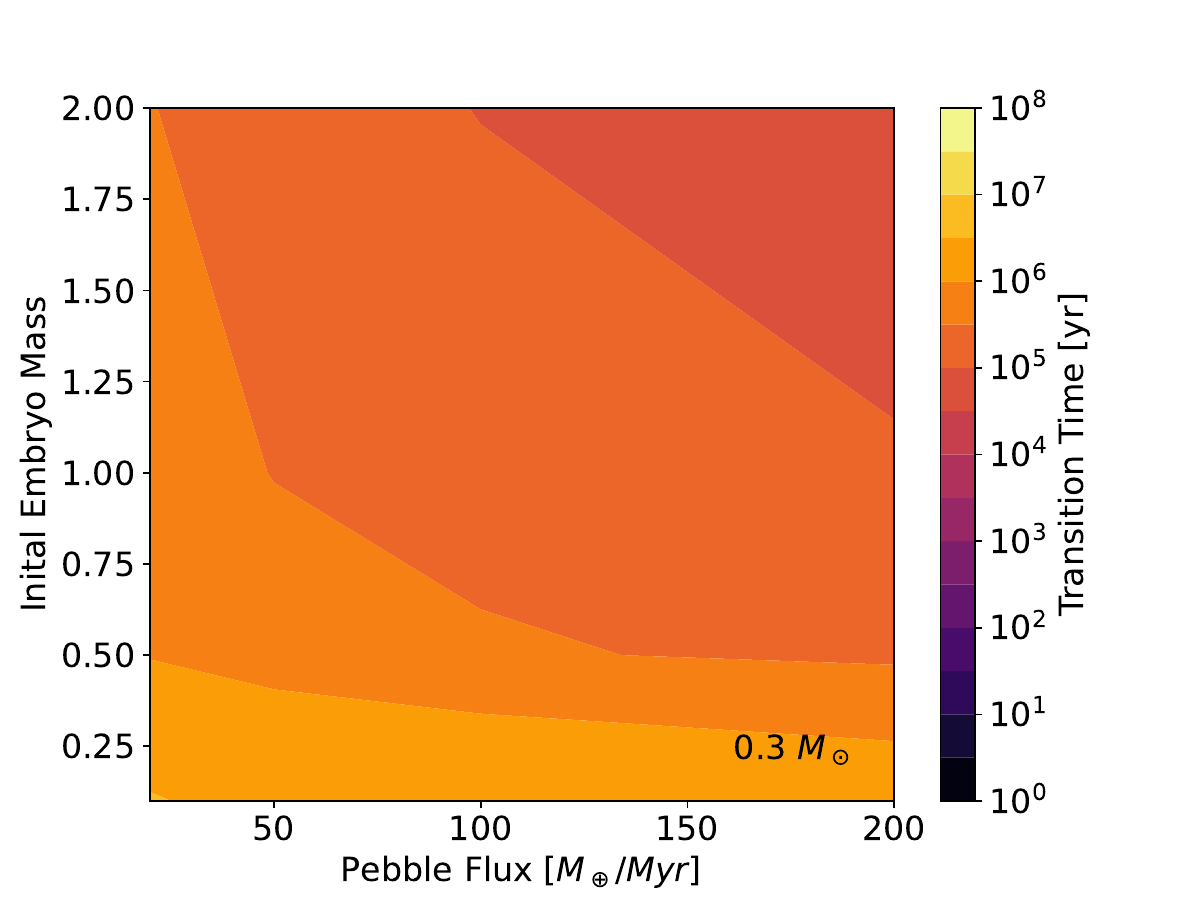}
    \end{minipage}
    \hfill
    \begin{minipage}[b]{0.45\textwidth}
        \centering
        \includegraphics[width=\textwidth]{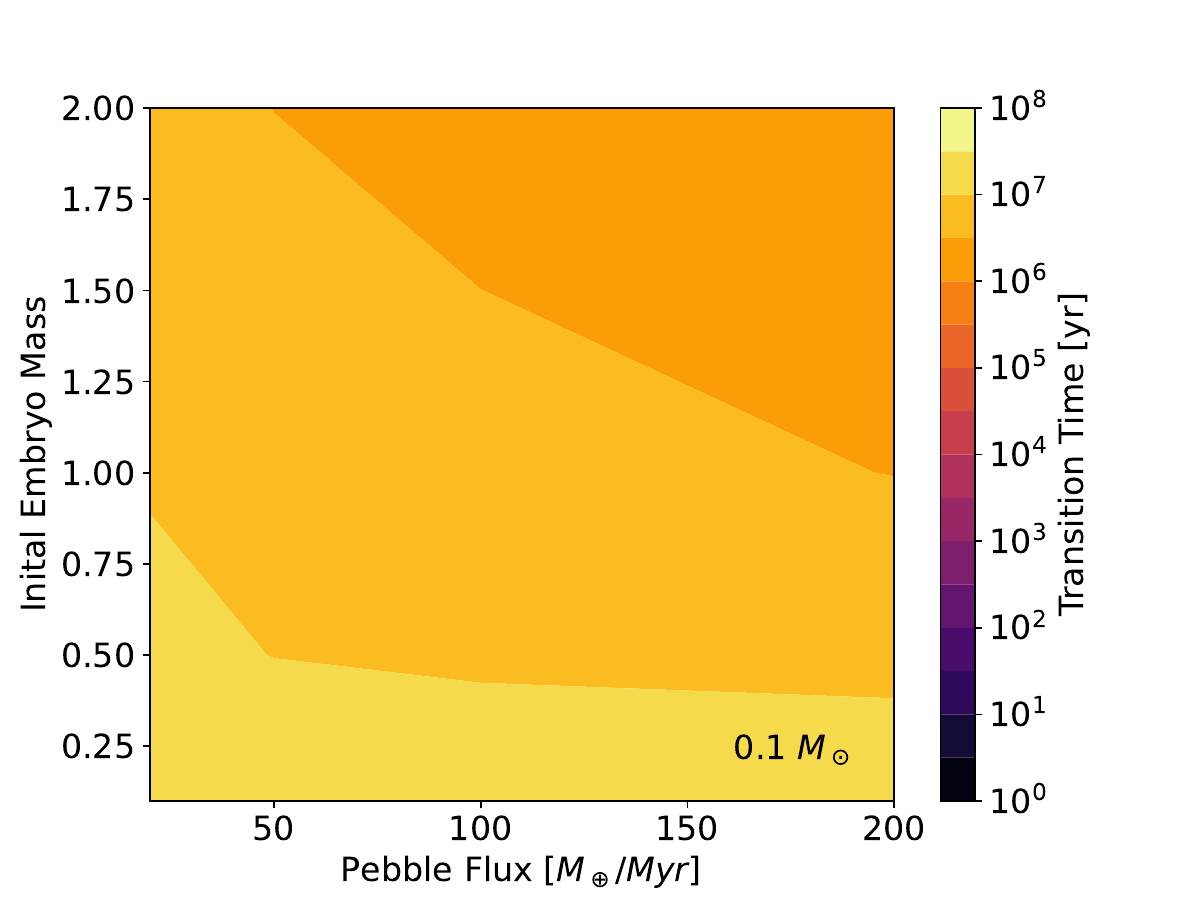}
    \end{minipage}
    
    \caption{Same as Figure \ref{fig:grid_plot} but for a $0.3 M_\odot$ (\textit{left}) and a $0.1 M_\odot$ (\textit{right}) mass star}
    \label{fig:grid_plots}
\end{figure*}

\end{appendix}

\end{document}